\documentclass[fleqn,usenatbib]{mnras}
\usepackage{newtxtext,newtxmath}

\usepackage[T1]{fontenc}

\usepackage{ae,aecompl}
\usepackage{comment}
\usepackage{microtype}
\usepackage{amsmath}
\usepackage{graphicx}
\usepackage{lscape}
\usepackage{color}
\usepackage{caption}
\usepackage{subcaption}

\graphicspath{{pics/}}
\newcommand {\bc}{\begin {center}}
\newcommand {\ec}{\end {center}}
\newcommand {\be}{\begin {equation}}
\newcommand {\ee}{\end {equation}}
\newcommand {\beq}{\begin {eqnarray}}
\newcommand {\eeq}{\end {eqnarray}}
\newcommand {\ergs}{{\rm erg\ \rm s$^{-1}$}}
\newcommand {\ergscm}{{\rm erg\ \rm s$^{-1}$\ \rm cm$^{-2}$}}

\definecolor{mypink1}{rgb}{0.858, 0.188, 0.478}

\def\flux{erg s$^{-1}$ cm$^{-2}$}

\def\sw{Swift\,J1808.4$-$1754}
\def\nustar{{\it NuSTAR}}
\def\swift{{\it Swift}}
\def\gaia{{\it Gaia}}

\title[Cyclotron line in the BeXRP Swift\,J1808.4$-$1754]
{Discovery of a pulse-phase-transient cyclotron line in the X-ray pulsar Swift\,J1808.4$-$1754 and identification of an optical companion}

\author[A.~Salganik et al.]
{Alexander~Salganik,$^{1,2}$\thanks{E-mail: alsalganik@gmail.com} 
Sergey~S.~Tsygankov,$^{3,2}$
Alexander~A.~Lutovinov,$^{2}$
Anlaug~A.~Djupvik,$^{4,5}$
\newauthor
Dmitri I. Karasev,$^{2}$ 
and Sergey~V.~Molkov$^{2}$
\\
$^1$Department of Astronomy, Saint Petersburg State University, Saint-Petersburg 198504, Russia\\
$^2$Space Research Institute of the Russian Academy of Sciences, Profsoyuznaya Str. 84/32, Moscow 117997, Russia\\
$^3$Department of Physics and Astronomy,  FI-20014 University of Turku, Finland\\
$^4$Nordic Optical Telescope, Apartado 474, 38700 Santa Cruz de La Palma, Santa Cruz de Tenerife, Spain
\\
$^5$Department of Physics and Astronomy, Aarhus University, Ny Munkegade 120, DK-8000 Aarhus C, Denmark
\\
}

\date{Accepted 2022 May 23. Received 2022 May 22; in original form 2022 April 20}

\pubyear{2022}

\begin{document}
\label{firstpage}
\pagerange{\pageref{firstpage}--\pageref{lastpage}}
\maketitle

\begin{abstract}
In this work, the temporal and spectral properties of the poorly studied X-ray pulsar Swift\,J1808.4$-$1754 were investigated in the 0.8-79~keV energy range based on the data from the \textit{NuSTAR} and \textit{Swift} observatories collected during the 2014 outburst. Strong pulsations with a period of $909.73\pm0.03$~s were detected in the source light curve, with the pulsed fraction demonstrating a nonmonotonic dependence on the energy with a local minimum around 17-22~keV. Phase lags in one of the pulse profile components, reaching the maximal value approximately at the same energy, were discovered. The pulse phase-averaged spectrum of the source has a power-law shape with an exponential cutoff at high energies, which is typical of X-ray pulsars. Pulse phase-resolved spectroscopy revealed the presence of a pulse phase-transient cyclotron absorption line at $\sim$21~keV, allowing us to estimate the neutron star magnetic field of $2.4\times10^{12}$~G. This makes Swift\,J1808.4$-$1754 a member of very small family of X-ray pulsars with a pulse-phase-transient cyclotron line in a narrow phase range. The data from the Nordic Optical Telescope allowed us to study the properties of the IR companion in the system and to conclude that most probably it is a Be-type star located at a distance of 5--8~kpc. 
\end{abstract}


\section{Introduction}
\label{intro}

Accreting X-ray pulsars (XRPs) in binary systems present a unique opportunity for studying extreme states of matter via timing and spectral analysis of their emission \citep[see e.g.,][]{2022arXiv220414185M}. 
The interaction of matter with the ultra-strong magnetic field ($\gtrsim 10^{12}$~G) of the neutron star (NS)  plays a key role in such systems. The most accurate and direct method to determine the XRP's magnetic field is to measure the energy of a cyclotron absorption feature in its energy spectrum \citep[see, e.g.,][and references therein]{Staubert2019}. This dictates the need to study XRP spectra in the widest possible energy range, which is provided by the {\it NuSTAR} observatory operating in the 3-79 keV range. 

A galactic transient X-ray source \sw\ was discovered with the \swift/BAT all-sky monitor during an outburst in May 2014  and localized with coordinates of RA$=18^{\rm h}08^{\rm m}26\fs4$, Dec. = $-17^{\circ}54\arcmin13\arcsec.7$ 
with an error radius of $3'$ at a 90\% confidence level \citep{Krimm2014}. A subsequent observation with the \swift/XRT telescope confirmed the source detection and improved its localization to RA$=18^{\rm h}08^{\rm m}25\fs14$, Dec. $= -17^{\circ} 53'49.2''$ with an error radius of $1.4''$ at a 90\% confidence level \citep{Krimm2014_2}. The XRT spectrum was well described by the absorbed power law model with the slope of $\Gamma$ = $0.93 \pm 0.29$ and absorption of $N_{\rm H} = 4.31^{+1.06}_{-0.93} \times 10^{22}$ cm$^{-2}$. The precise {\it Swift}/XRT localisation allowed \citet{Krimm2014_2} to propose the star 2MASS 18082507$-$1753482 as a possible companion of \sw. This star was detected neither in the \swift/UVOT b-band nor the DSS data.

Based on the ratio of \swift/BAT and \swift/XRT count rates \citet{Krimm2014_2} suggested that the source is more likely to be a neutron star. Further observations with \swift/XRT have revealed a strong flux modulation with a period of $924\pm3$ seconds, which was interpreted as a periods of the NS rotation \citep{Tomsick2014}.  
Based on the fact that the source is seen in the near-IR but not in the optical, these authors proposed the system to be a high-mass X-ray binary (HMXB). The \nustar\ observation performed at the declining phase of the outburst provided more accurate estimation of the \sw\ spin period of 909.73(11) seconds \citep{Bachetti2014} with the discrepancy between two values explained by the non-sinusoidal pulse profile and the sparse sampling of the pulse in the short \swift/XRT observations.

In this paper, we present results of the first detailed study of the XRP \sw\ in a wide energy range of 0.8-79 keV, carried out using data from the Neil Gehrels \swift\ and \nustar\ observatories. Additionally, the class of the counterpart star 2MASS~18082507$-$1753482 was determined based on the near-IR spectroscopic data of the Nordic Optical Telescope (NOT).

\section{Data analysis}
\label{sec:data}

\begin{figure}
    \centering
    \includegraphics[width=0.95\columnwidth]{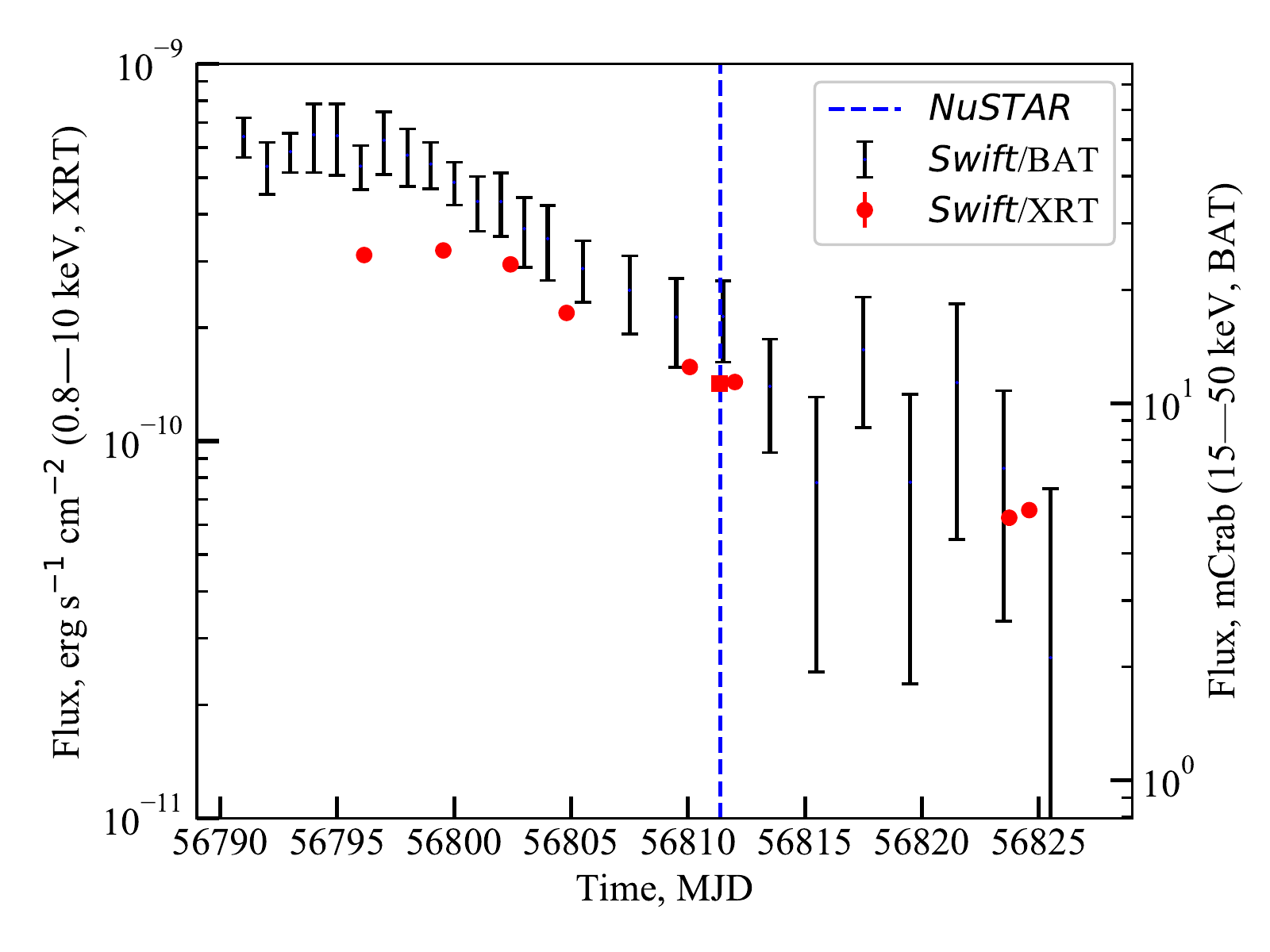}
 	\caption{The light curve of \sw\ according to the {\it Swift}/BAT and {\it Swift}/XRT telescopes in the 15--50 keV and 0.8-10 keV  energy ranges, respectively. The moment of the {\it NuSTAR} observation is shown with the vertical dashed line. BAT data are one-day and two-day binned before and after MJD 56805, respectively. For a conversion of the BAT light curve into mCrab units, the following relation was applied: 1 mCrab =  0.00022 counts~cm$^{-2}$~s$^{-1}$. The square marks the observation used in the spectral analysis (see Sec.~\ref{sec:spectrum}). Fluxes for both instruments are given in the absolute values.}
 	\label{fig:lightcurve}
\end{figure}

\subsection{\nustar\ observatory}

The \nustar\ observatory has two identical co-aligned X-ray telescopes focusing X-ray photons onto two Focal Plane Modules FPMA and FPMB and operates in the 3-79 keV energy range \citep{Harrison2013}, providing a high sensitivity in hard X-rays.
\nustar\ observed \sw\ during the decay of the 2014 outburst (June 3, ObsID 80002014002; Fig.~\ref{fig:lightcurve}) with an effective exposure of 42\,ks.
The source events for the analysis were selected from a circular region with the radius of 40\arcsec. The radius of the background region is 120\arcsec\ located far enough from the source. Sizes of the regions were selected to optimize the signal-to-noise ratio at high energies.

Data reduction was performed with the standard \nustar\ Data Analysis Software ({\sc nustardas}) v2.0.0 provided under {\sc heasoft} v6.29 with the {\sc CALDB} 4.9.6 calibration files package in accordance with the data analysis manual.\footnote{\url{https://heasarc.gsfc.nasa.gov/docs/nustar/analysis/nustar_swguide.pdf}}
Barycentric correction with the {\sc barycorr} utility was applied to the background-subtracted FPMA and FPMB light curves of the source. The light curves for two modules were summed using the {\sc lcmath} utility in order to increase count statistics.
Energy spectra of the source were binned in accordance with the optimal binning algorithm proposed by \citet{Kaastra2016}, which is implemented in the {\sc ftgrouppha} utility from the {\sc FTOOLS} package. 

\subsection{\swift\ observatory}

To broaden the energy range available for the spectral analysis of \sw\ we utilized the {\it Swift}/XRT telescope observation (ObsID 00080746001) performed in the Windowed Timing mode (WT) with the exposure of 6.9 ks simultaneously with the \nustar\ observation. 
Also, to study the evolution of the source flux during the outburst in the soft (0.8-10~keV) X-ray band, observations of the {\it Swift}/XRT telescope (ObsID 00033287001, 00033287002, 00033287003, 00033287004, 00033287005, 00080746001, 00080746002, 00033287009, 00033287010) were additionally used (see Fig.~\ref{fig:lightcurve}). The resulting light curve including the \textit{Swift}/BAT data\footnote{\url{https://swift.gsfc.nasa.gov/results/transients/weak/SWIFTJ1808.4-1754/}} is plotted for the visual comparison. The \textit{Swift}/XRT light curve was obtained by fitting spectra in each observation with a simple absorbed power-law model with the fixed column density $N_{\rm H} = 4.0\times10^{22}$ cm$^{-2}$, which was determined from a broad band spectrum of the source (see Sect.\ref{sec:spectrum}). To convert a BAT light curve to the mCrab units the Crab count rate of 0.22~counts~cm$^{-2}$~s$^{-1}$ was assumed \citep{Romano2014}.

To investigate properties of the source in the quiescent state, that was never done before, we triggered several TOO observations with the \swift/XRT telescope. The total exposure of 6 ks was obtained by stacking four observations (ObsID 00033287011, 00033287012, 00033287013, 00033287014) performed from 22 August 2021 to 4 September 2021. 

The {\it Swift}/XRT spectra of \sw\ during the outburst were obtained using the online service provided by the UK {\it Swift} Science Data Center \citep{Evans2009}.\footnote{\url{https://www.swift.ac.uk/user\_objects/}} The resulting spectra were grouped to have at least 1 count per energy channel. 

The resulting broadband spectrum of the source from all instruments was fitted with different models using W-statistics\footnote{\url{https://heasarc.gsfc.nasa.gov/xanadu/xspec/manual/XSappendixStatistics.html}} \citep{Wachter1979} in the {\sc XSPEC} 12.11.1 package \citep{Arnaud1996}. All error are given at $1\sigma$ confidence level if not specified otherwise.

\begin{figure}
    \centering
    \begin{subfigure}[b]{0.79\columnwidth}
        \includegraphics[height=8cm,keepaspectratio]{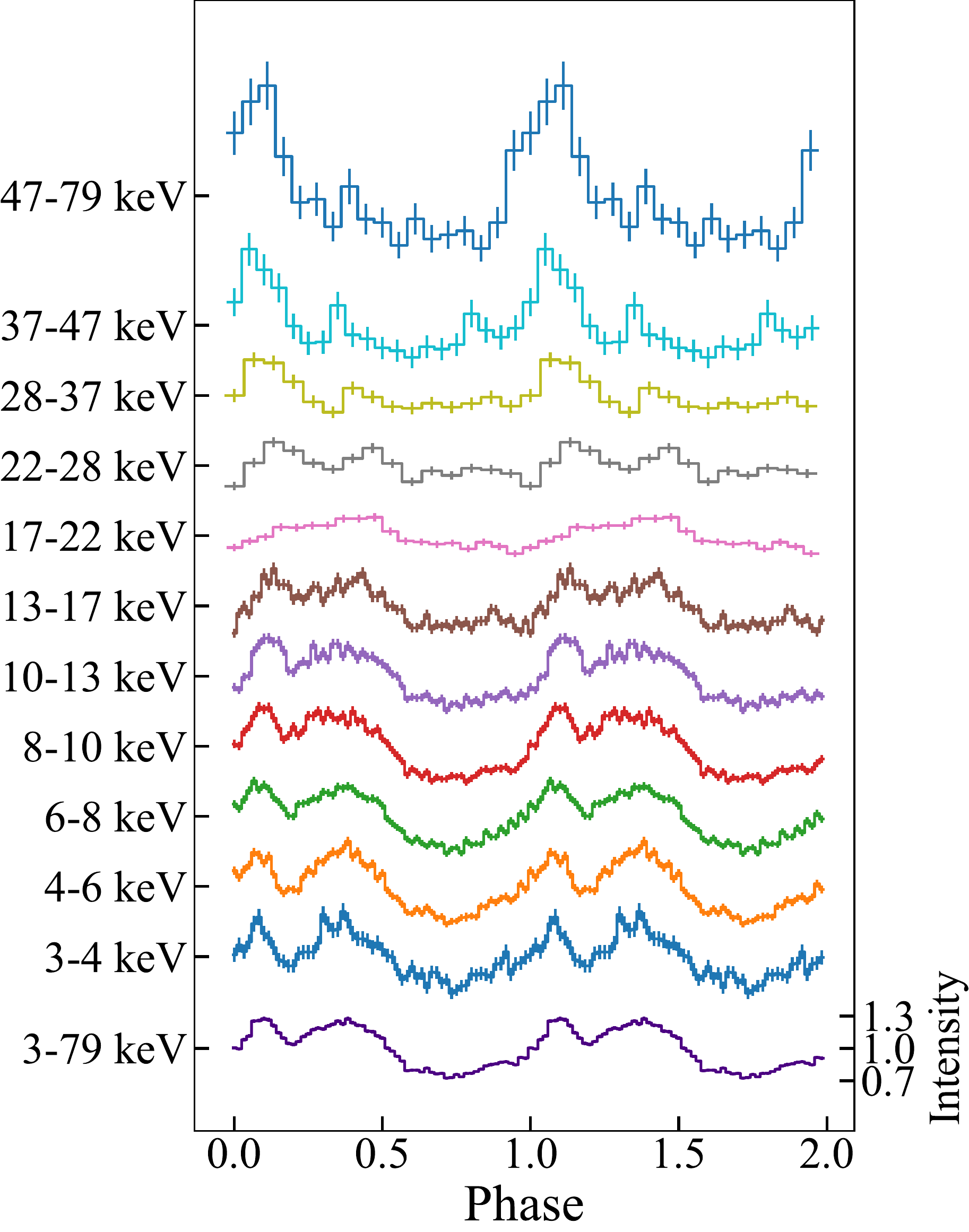}
        \caption{Pulse profiles.}
        \label{fig:profiles}
    \end{subfigure}
    \begin{subfigure}[b]{0.2\columnwidth}
        \includegraphics[height=8cm,keepaspectratio]{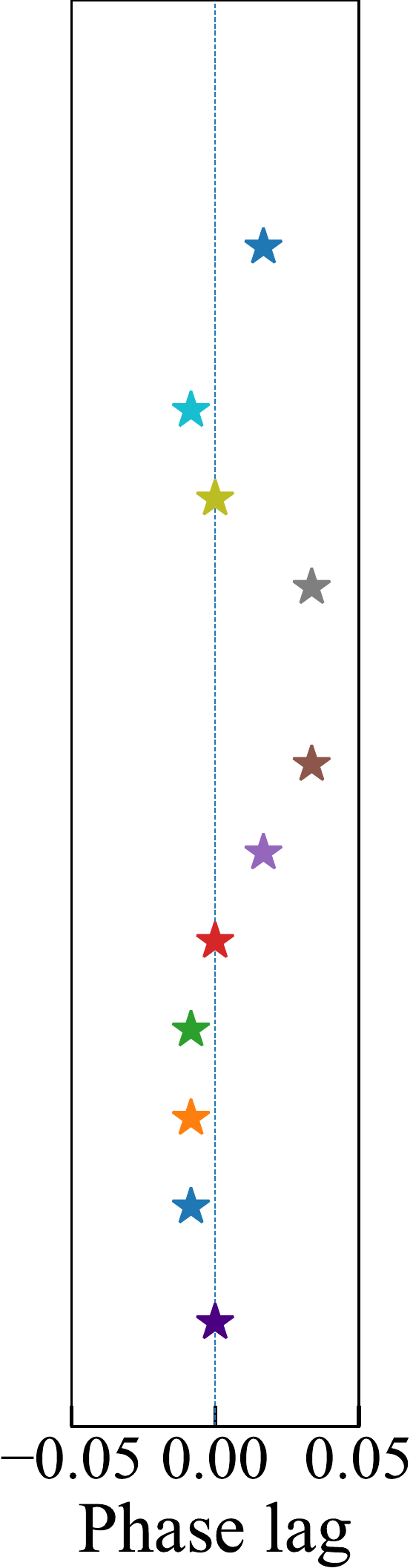}
        \caption{Phase lags.}
        \label{fig:lags}
    \end{subfigure}\\
    \begin{subfigure}[b]{0.99\columnwidth}
        \includegraphics[height=4.8cm,keepaspectratio]{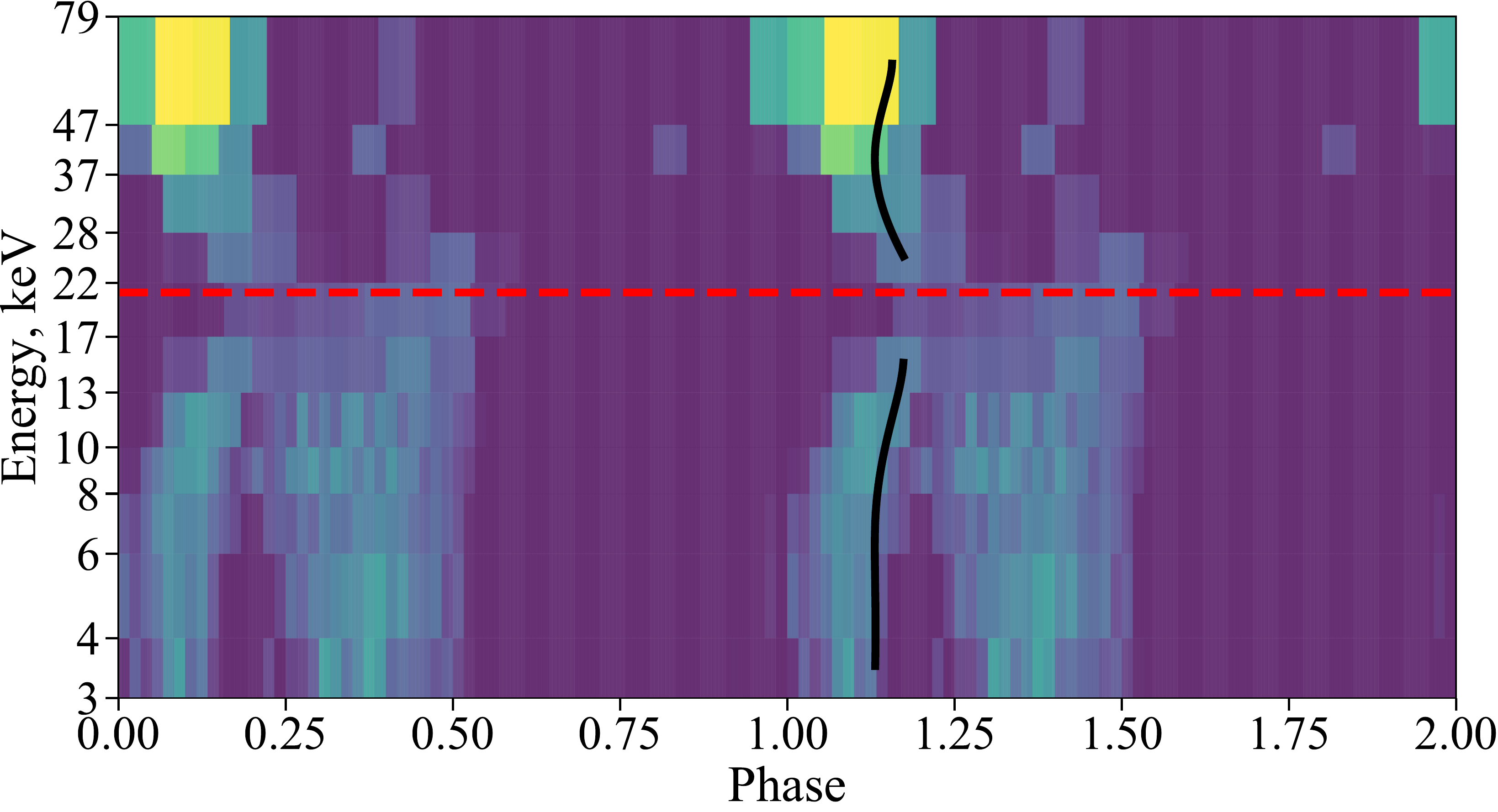}
        \caption{2D distribution of pulse profiles.}
        \label{fig:2D}
    \end{subfigure}
    \caption{Figure~\ref{fig:profiles}: \sw\ pulse profiles in different energy bands. Figure~\ref{fig:lags}: corresponding phase lags (see details in Sec.~\ref{sec:timing}). 
    \\\hspace{\textwidth}
    Figure~\ref{fig:2D}: 2D Energy-Phase distribution of the \sw\ pulse profile, the energy scale is logarithmic. The phase lags spline interpolation is overplotted with the black line for the visual comparison. The 21~keV cyclotron line energy $E_{\rm cyc}$ is shown by the red dashed line (see Sec~\ref{sec:phase_spec}).  
    }
    \label{fig:summary_profiles}
\end{figure}

\subsection{Nordic Optical Telescope}

The near-infrared Camera and Spectrograph, NOTCam \citep{Abbott2000}, with its Hawaii-1 HgCdTe detector was used at the 2.56 m {\it NOT} \citep{Djupvik2010} on 2021 June 20 to take $K$-band spectra of the near-IR counterpart of \sw. We used the WF-camera (0.234$\arcsec$/pix) with Grism $\#1$,
a 0.6$\arcsec$ wide slit, and the MKO $K$-band filter (\#208) as order sorter,  giving a resolving power of $\lambda/\Delta \lambda$ = 2100. The observations were obtained with multiple AB mode dithering along the slit to allow for sky subtraction. Each single spectrum of 300 s was a result of 10 ramp-sampling readouts every 30 seconds. Due to a wave of bad seeing during these observations many spectra had to be discarded and in the end we used a total of 8 individual spectra, i.e. a total exposure time of 2400 seconds. Immediately before the target we observed the A0\,V star HD161822 as a telluric standard. Details on the observations and data reductions are described in \citet{Salganik2022}. 

\section{Results}

\subsection{Timing analysis}
\label{sec:timing}

For the detailed timing analysis we used only \nustar\ data, providing a long uninterrupted observation of \sw. Note, that parameters of the binary system containing \sw\ are still unknown, therefore all following results of the timing analysis are uncorrected for the orbital motion of the neutron star. A standard epoch folding technique, which is implemented in the {\sc efsearch} tool from the {\sc ftools} package, revealed strong pulsations of the flux with a period of $P_{\rm spin} = 909.73 \pm 0.03$ s, what is consistent with the period obtained by \citet{Bachetti2014}. 
An uncertainty for the period  was estimated from distribution of pulse periods obtained from a large number of simulated light curves following the procedure described by \citet{Boldin2013}.
A sufficiently high count statistics provided by the {\it NuSTAR} observatory allowed us to study a dependence of the \sw\ pulse profile on the energy range (see Fig.~\ref{fig:profiles}). 
As can be seen from the figure, at energies $\leq 30$ keV the profile is sinusoidal with a main peak at 0.0-0.5 phases consisting of two humps. At energies $\lesssim 17$~keV, the humps intensities remain relatively constant, while in the energy range 17-22 keV the pulse profile becomes much flatter and the left hump almost disappears. At energies $\gtrsim$ 22~keV, the intensity of the left hump relative to the right one begins to increase until the right peak disappears at energies of 47-79 keV.

\begin{figure}
    \centering
    \includegraphics[width=0.95\columnwidth]{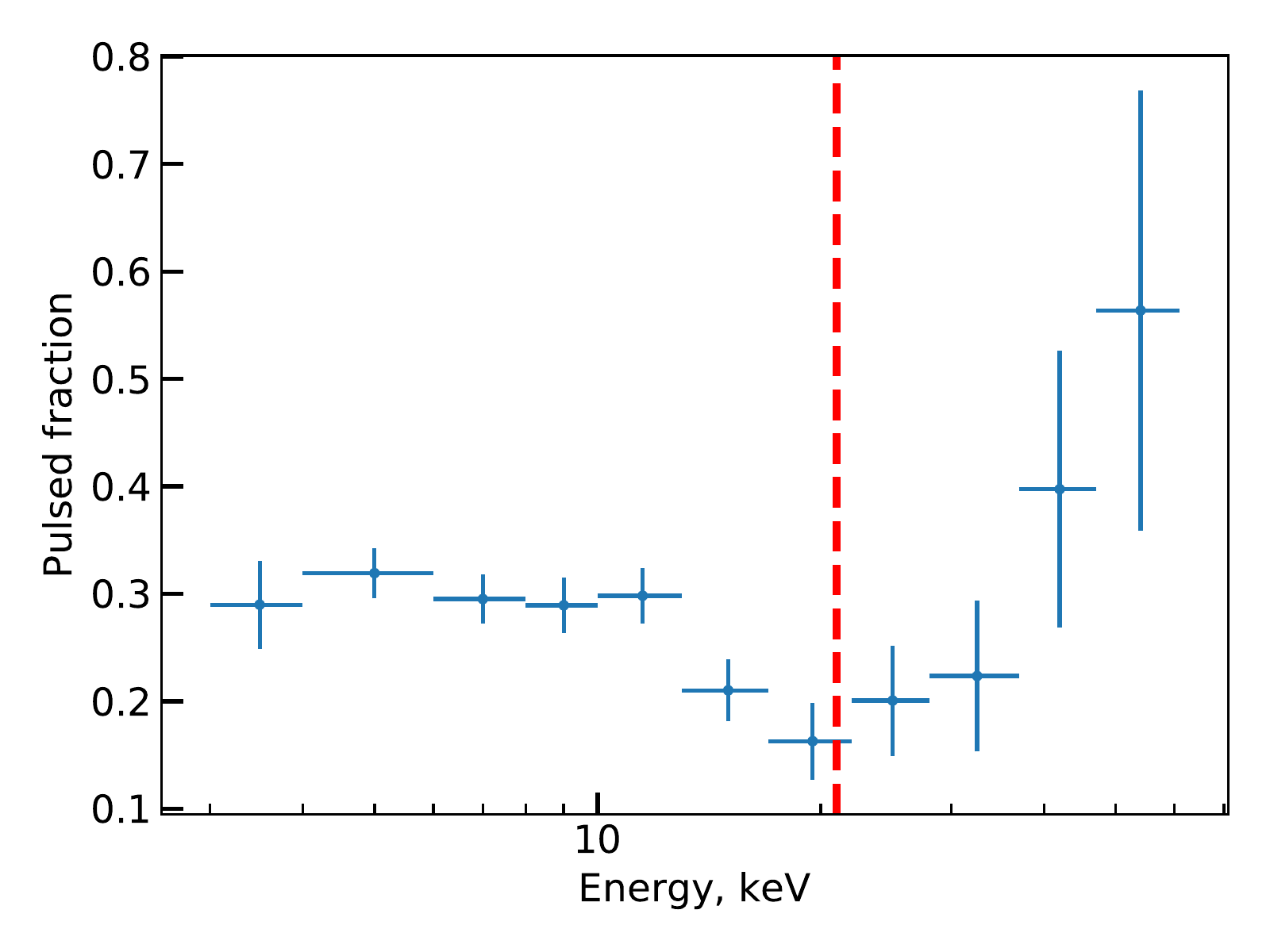}
	\caption{Dependence of  the \sw\ pulsed fraction on the energy based on the \nustar\ data. The 21~keV cyclotron line energy $E_{\rm cyc}$ is shown by the red dashed line (see Sec~\ref{sec:phase_spec}). }
	\label{fig:pulsed}
\end{figure}

To study changes of the pulsation amplitude in more details, we used energy-resolved folded light curves of the pulsar to derive the dependence of the pulsed fraction (PF), defined as $({\max C - \min C})/({\max C + \min C})$, where min$C$ and max$C$ are the minimum and maximum count rates in the pulse profile, respectively, on the energy (see Fig. ~\ref{fig:pulsed}). Pulse profiles used in the PF calculation have 15 phase bins in each energy band. As can be seen from the figure, the pulsed fraction increases non-monotonously: at low energies it is practically independent of the energy, staying around 30\%,  then it drops down to $\sim$15\% at 17-22 keV and increases at higher energies. The latter one is typical for most of XRPs \citep{LutovinovTsygankov2009}, but the PF drop around 17-22 keV looks quite intriguing. 

The dependence of the \sw\ pulse profile on the energy shows also a presence of the hard phase lag (see Fig.~\ref{fig:lags} and Fig.~\ref{fig:2D}). To reveal it, the pulse profile in each energy range was binned to have 60 phase bins and a cross-correlation of the main peak (located at phases of 0.0-0.28) in a given energy range with the one in the total 3-79 keV range was performed. The cross-correlation analysis was made in this narrow phase interval since the lags are most prominent there. In the 17-22 keV range, the main peak is almost absent, that doesn't allow to calculate the phase lags correctly. 
The disappearance of the main peak at these energies can be clearly seen in the two-dimensional distribution of the normalized intensity of the pulse profile as a function of the energy (Fig.~\ref{fig:2D}) and in the pulsed fraction dependence on the energy, where a significant decrease is observed (Fig.~\ref{fig:pulsed}).

\subsection{Pulse phase-averaged spectral analysis}
\label{sec:spectrum}

\begin{figure}
\centering
\includegraphics[width=0.95\columnwidth]{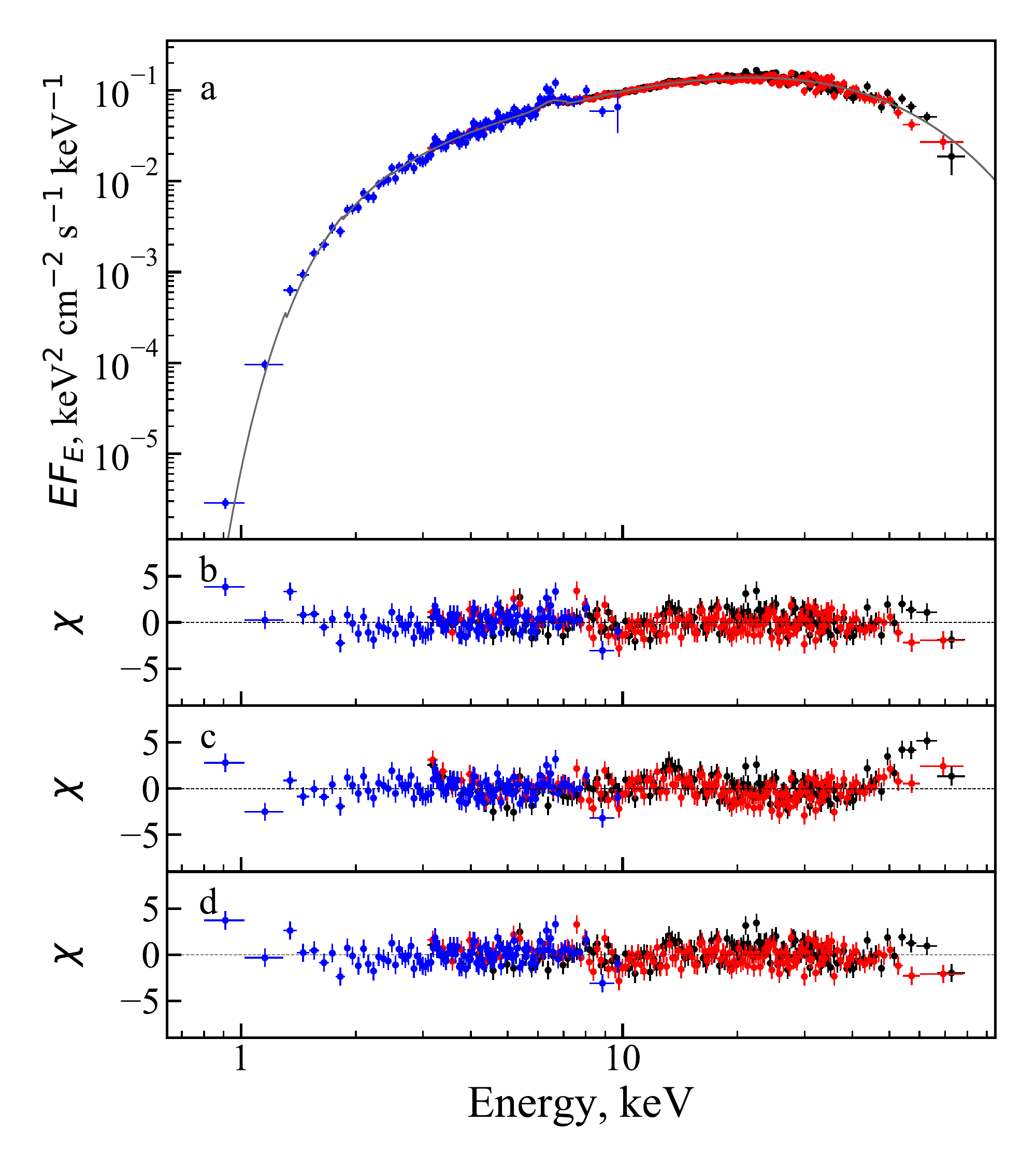}
\caption{Unfolded energy spectrum of \sw\ and its approximation with the model {\sc tbabs $\times$ (cutoffpl + gauss)} (solid grey line in panel a). Red and black dots show data from the FPMA and FMPB telescopes of the {\it NuSTAR} observatory, respectively; blue ones are for the {\it Swift}/XRT telescope. The bottom three panels show the deviations of the data from different continuum models: {\sc cutoffpl} (panel b), {\sc compTT} (panel c), {\sc pow $\times$ highecut} (panel d).} 
\label{pic:avg_spectrum}
\end{figure}

Fig.~\ref{pic:avg_spectrum} demonstrates the phase-averaged energy spectrum of the XRP \sw\ according to the {\it NuSTAR} (ObsID 80002014002) and {\it Swift}/XRT (ObsID 00080746001) data, obtained during observations on June 3 and 2, 2014, respectively. A simultaneous fit of these data cover energies from 0.8 to 79 keV. 

The source spectrum has a typical shape for XRPs \citep[e.g.][]{Filippova2005, Coburn2002}. To fit the spectrum, we applied several phenomenological models broadly used to approximate spectra of XRPs. As a result we found that the source spectrum is best described by an exponential cutoff power law model {\sc cutoffpl} in {\sc xspec} (W-stat/dof = 1217/1187). To account for the possible systematic uncertainties in the calibration of {\it NuSTAR}/FPMA, {\it NuSTAR}/FPMB and {\it Swift}/XRT, we used a cross-calibration factor (the multiplicative component {\sc const} in the model) which was frozen to 1.0 for FPMA and kept free for FPMB and XRT. 
A photoabsorption was taken into account using the {\sc tbabs} component with the abundance adopted from \citet{Wilms2000}. The spectrum has the strong iron fluorescent $K_\alpha$ line at 6.4 keV, which was described with a Gaussian profile ({\sc gauss} model). No additional emission or absorption features were required to fit the pulse phase-resolved spectrum of \sw.

An exponential cutoff power law model with the cutoff at high energies  {\sc pow $\times$ highecut} describes the spectrum well to the same extent (W-stat/dof = 1209/1185). Note, that the multiplicative {\sc gabs} component with the energy $E_{\rm smoothgabs}$ = $E_{\rm cut}$ and $\sigma_{\rm smoothgabs}$ = 0.1 $E_{\rm cut}$ was added to the {\sc pow $\times$ highecut} model in order to smooth out the discontinuity resulting in an artificial absorption-like feature \citep[see, e.g.,][for details]{Coburn2002}. Nevertheless, since the {\sc pow $\times$ highecut} model is more complex, it was decided to use {\sc cutoffpl}, which has less free parameters (see Table \ref{table:spec_params}). Finally, we also tested a Comptonized radiation model ({\sc compTT}) from \citet{Titarchuk1994}, which is usually considered as a good approximation for the XRPs spectra. However, {\sc cutoffpl} statistic and approximation quality is much better than {\sc compTT} (see Table \ref{table:spec_params}).

\begin{table*}
\caption{Best-fit spectral parameters of \sw\ for different continuum models obtained from the pulse phase-averaged spectral analysis.}
\begin{center}
\begin{tabular}{lccc} 
 \hline
 Parameter &{\sc cutoffpl} &  {\sc pow $\times$ highecut}
 & {\sc compTT} \\
  \hline
 ${\rm const}_{\rm FPMA}$ & 1.000 (frozen)  &  1.000 (frozen) &  1.000 (frozen)  \\ 
 ${\rm const}_{\rm FPMB}$ & $0.980 \pm 0.004$ &  $0.980 \pm 0.004$ &  $0.980 \pm 0.004$ \\ 
 ${\rm const}_{\rm XRT}$ &  $0.928 \pm 0.015$  &  $0.930 \pm 0.015$ &  $0.942 \pm 0.015$  \\
 $N_{\rm H}$, $10^{22}$~cm$^{-2}$ & $4.0\pm0.2$ & $3.6^{+0.2}_{-0.1}$ & $1.5\pm0.1$ \\
 $E_{\rm cut}$,~keV &  &$4.1^{+0.2}_{-0.3}$ &   \\
 $E_{\rm fold}$,~keV & $15.7\pm0.2$ & $16.0\pm0.3$ & \\ 
 $\Gamma$ & $0.62\pm0.01$ & $0.63\pm0.02$ &  \\ 
 $T_0$,~keV & &  & $1.28\pm0.02$\\ 
 $T$,~keV & & &$7.71^{+0.08}_{-0.07}$   \\ 
 $\tau$ & & &$4.92\pm0.05$   \\
 geometry $\beta$ & & & 1.0 (frozen)   \\
 ${\rm Norm}_{\rm continuum}$, ph~keV$^{-1}$~s$^{-1}$~cm$^{-2}$  & $(7.9\pm0.2)\times10^{-3}$ &  $(6.3\pm0.2)\times10^{-3}$ & $(4.0\pm0.04)\times10^{-3}$   \\
 
 $E_{\rm smoothgabs}$,~keV &  &$4.1$ (=$E_{\rm cut}$) & \\ 
 $\sigma_{\rm smoothgabs}$,~keV &  &$0.4$ (=$0.1 \times E_{\rm cut}$) & \\
 $\tau_{\rm smoothgabs}$ &  &$0.024\pm0.016$ & \\
 $E_{\rm gauss}$,~keV & $6.46^{+0.03}_{-0.04}$ &  $6.47 \pm 0.03$ & $6.49 \pm 0.03$ \\
 $\sigma_{\rm gauss}$,~keV & $0.38^{+0.08}_{-0.06}$ &  $0.32^{+0.06}_{-0.05}$ & $0.21 \pm 0.04$ \\
 ${\rm Norm}_{\rm gauss}$, ph~s$^{-1}$~cm$^{-2}$  & $(2.6 \pm 0.3)\times 10^{-4}$& $(2.3 \pm 0.3 )\times 10^{-4}$ & $(1.5 \pm 0.2)\times 10^{-4}$ \\
 Flux$_{0.8-79 \rm keV}$, \flux & $(4.80\pm0.03)\times10^{-10}$ & $(4.80\pm0.03)\times10^{-10}$ & $(4.55\pm0.02)\times10^{-10}$ \\
  W-statistic/d.o.f. & 1217/1187 & 1209/1185 & 1290/1186 \\
 \hline
\end{tabular}
\label{table:spec_params}
\end{center}
\end{table*}

Interestingly, the neutral hydrogen column density $N_{\rm H}$ for the {\sc compTT} model differs significantly from its values for the {\sc pow $\times$ highecut} and {\sc cutoffpl}. This indicates that the $N_{\rm H}$ value is highly dependent on the chosen model and has a large systematic uncertainty. The value of the column density of the neutral hydrogen obtained from the model with the {\sc cutoffpl} continuum turns out to be $\sim$ 4-5 times higher than the Galactic value in the direction to the source $0.9\times10^{22}~{\rm atoms}~{\rm cm}^{-2}$ \citep{HI4PI2016}, that may be related to the internal absorption in the system.

\begin{figure}
    \centering
 	\includegraphics[width=0.95\columnwidth]{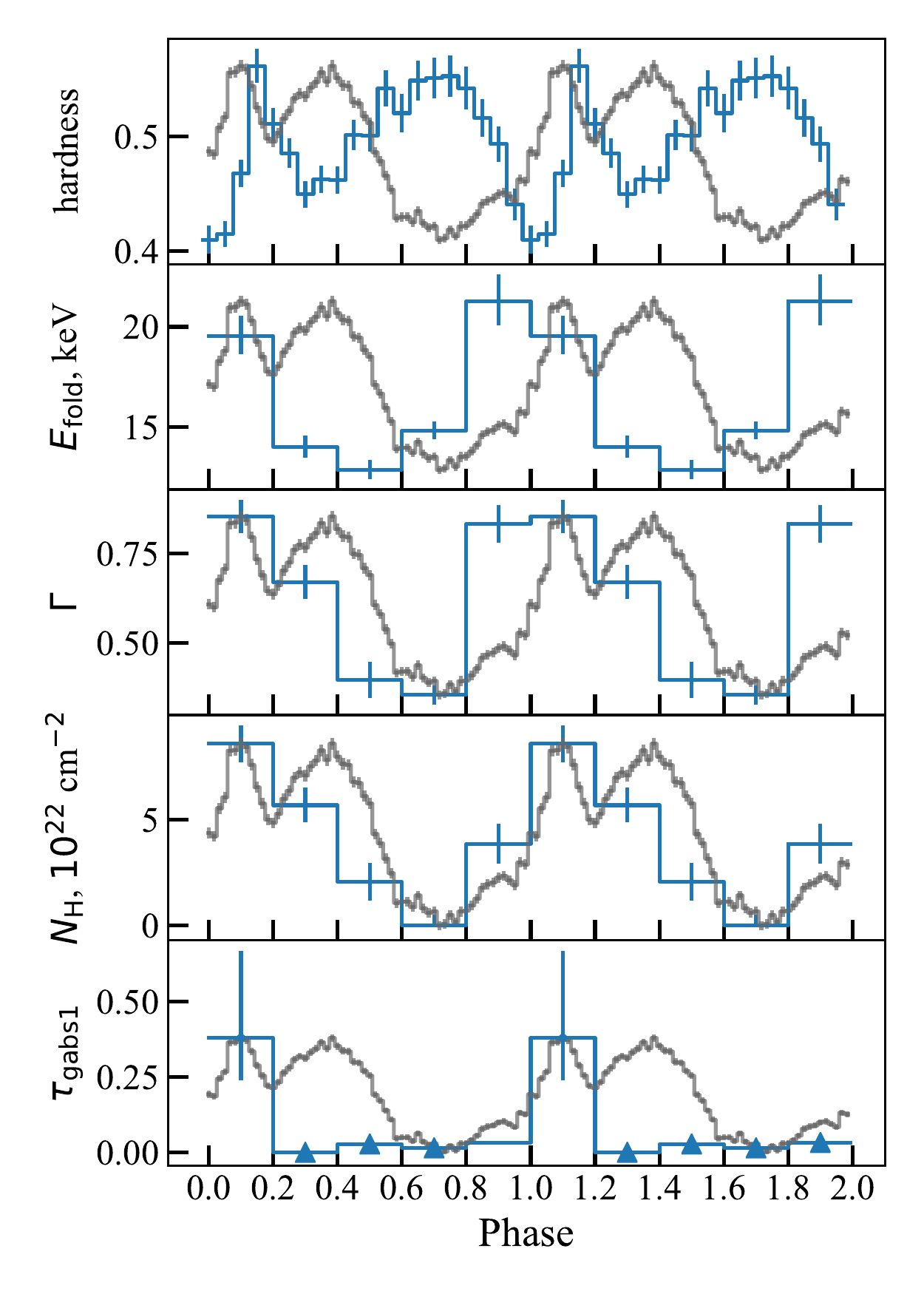}
 	\caption{The hardness ratio of the \sw\ emission and the evolution of the spectral model parameters as a function of the spin phase. The averaged pulse profile in a wide energy range is superimposed in gray for the visual comparison. The hardness is defined as the ratio of unnormalized pulse profiles in the 10-20 keV / 3-10 keV energy ranges. The triangles represent the 3$\sigma$ upper limits for 21~keV CRSF in the phase-resolved spectra.}
	\label{fig:spec_params}
\end{figure}

\subsection{Pulse phase-resolved spectroscopy}
\label{sec:phase_spec}

\begin{figure}
    \centering
 	\includegraphics[width=0.98\columnwidth]{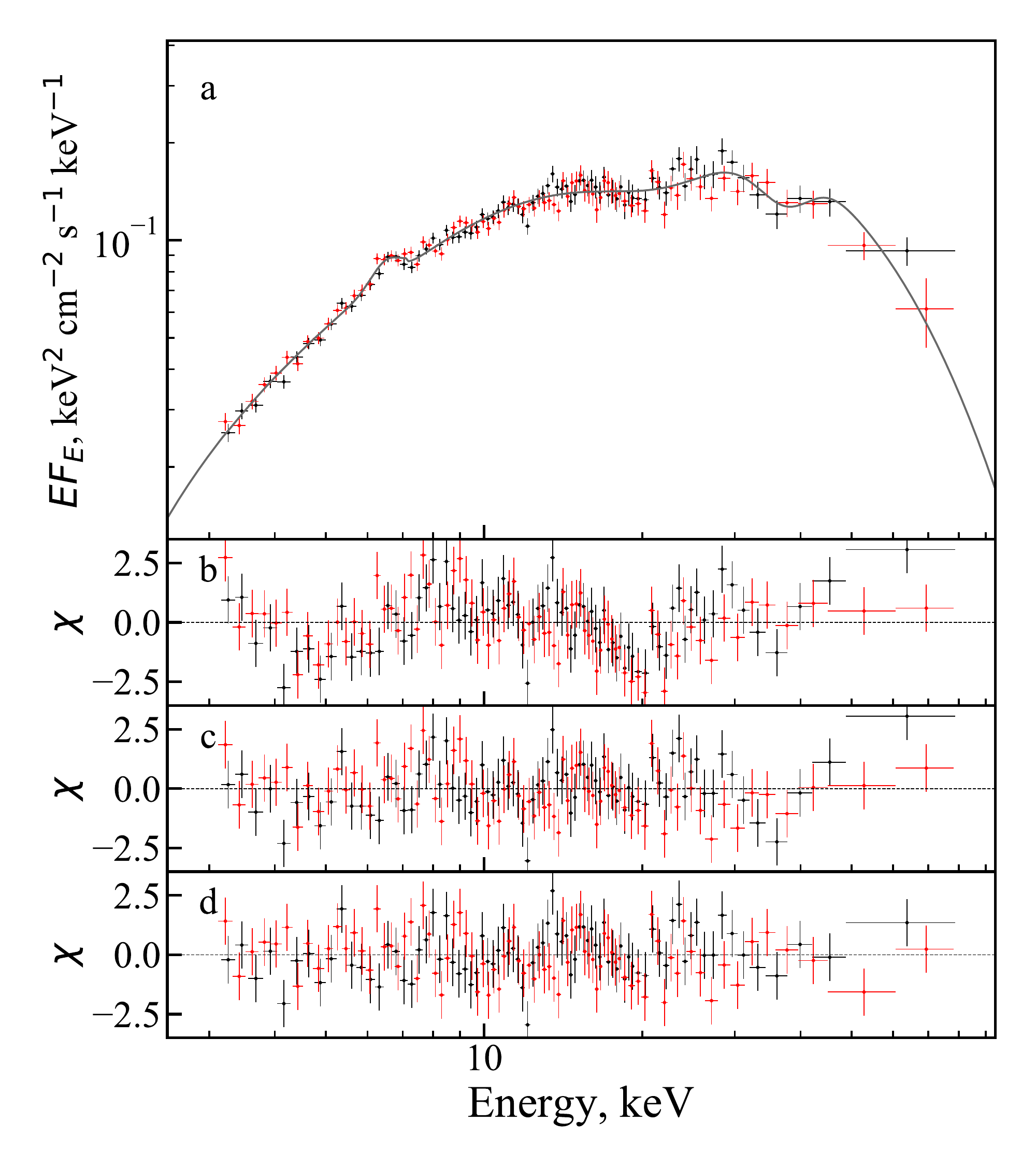}
 	\caption{Energy spectrum of \sw\ at the pulse phases 0.0–0.2 for the \nustar\ observation and its approximation with the model {\sc tbabs $\times$ (cutoffpl $\times$ gabs $\times$ gabs + gauss)} (solid grey line in panel a). The data from the FPMA and FPMB modules are shown by red and black dots, respectively. 
 	 The bottom three panels show the deviations of the data from models without both {\sc gabs} components (panel b), without only 37~keV harmonic component (panel c), with both components included to the model (panel d).}
	\label{fig:cyclotron}
\end{figure}

High quality of the \nustar\ data allowed us to carry out a pulse phase-resolved spectroscopy and study the evolution of spectral parameters as a function of the rotation phase of the NS. Data were divided into 5 evenly distributed phases (0.0-0.2, 0.2-0.4, ..., 0.8-1.0, see Fig.~\ref{fig:spec_params}).
To approximate the phase spectra, we used the best-fit model {\sc tbabs $\times$ (cutoffpl + gau)} as for the phase-averaged spectrum (energy and width of the iron line were frozen at values $E_{\rm gauss} = 6.46$~keV, $\sigma_{\rm gauss} = 0.39$~keV, obtained from the phase-averaged spectrum). 

Fig.~\ref{fig:spec_params} shows an evolution of the spectral parameters $E_{\rm fold}$, $\Gamma$, $N_{\rm H}$ and hardness ratio with the rotational phase. It is nicely seen that all three parameters exhibit strong sinusoidal variations with a large amplitude and show a strong correlation. The hardness ratio exhibits a two-peaks structure with maxima at phases of 0.0-0.3 (narrow one) and 0.4-1.0 (broad one), and shows a significant anti-correlation with the pulse profile at phases of 0.2-0.9.

The spectrum in the 0.0-0.2 phases is especially different from the other ones. It shows strong deviations from the continuum model around 20-21 keV (see panel b in Fig.~\ref{fig:cyclotron}) and can be described by the above mentioned model with W-stat/dof = 487/361. It is known that in some XRPs with cyclotron resonant scattering feature (CRSF), its strength may vary significantly over the pulse phase or even appear only in a narrow range of phases \citep[see, e.g.,][]{Kreykenbohm2002, Doroshenko2017, Molkov2019, Molkov2021}. Therefore, the observed deviations from the best-fit continuum model we interpret as a possible pulse-phase transient CRSF. An inclusion of the corresponding {\sc gabs} component to the model significantly improves the W-stat/dof to 410/358. Even after that some small deviations from the model can still be seen around 37 keV (see panel c in Fig.~\ref{fig:cyclotron}). An addition of the second {\sc gabs} component results in an improvement in W-stat/dof to 391/355. This second line can be considered as the first harmonic of the fundamental cyclotron feature at $\simeq21$~keV. The best-fit parameters of the spectral models containing different number of CRSFs are presented in Table~\ref{table:cyclotron}.

To assess the significance of the detected CRSFs, we performed Monte Carlo simulations  using the {\sc simftest}\footnote{\url{https://heasarc.gsfc.nasa.gov/xanadu/xspec/manual/node125.html}} procedure from the {\sc XSPEC} package, providing the value of the statistical difference (difference of W-stat values) between the models with and without a specific CRSF component. First, to calculate the significance of the $\simeq21$~keV CRSF, we made $10^{4}$ simulations with the {\sc gabs} component added to the continuum model {\sc tbabs $\times$ (cutoffpl + gauss)}. At the next step, we made additional $10^{4}$ simulations with the second {\sc gabs} component at $\simeq37$~keV added to the model already including the $\simeq21$ keV feature.

To determine a significance level of the discovered absorption features, the corresponding histograms of the statistical difference were approximated by the $\chi^2$ distribution with three degrees of freedom (number of {\sc gabs} model parameters). Note, that the chi-squared distribution does not accurately describe the obtained distributions, but it is suitable as an estimate. Thus, we get a probability of a false detection with the addition of the $\simeq21$ keV cyclotron line equal to $10^{-16}$ and a probability of a false detection of the subsequent addition of the $\simeq37$ keV line equal to $3\times10^{-4}$. This indicates a highly significant CRSF detection at $\simeq21$ keV and a hint to the presence of its first harmonics  around 37 keV at the pulse phases 0.0-0.2 (see Table~\ref{table:cyclotron}). Fig.~\ref{fig:spec_params} shows that there is significant detection of the cyclotron line only in the narrow 0.0-0.2 phase range with the line depth 5-67 times larger in comparison to the other phases.

\begin{table*}
\caption{The best-fit spectral parameters for the pulse phase interval 0.0-0.2 for different models (see text).}
\begin{center}
\begin{tabular}{lccc} 
 \hline
 Parameter &{\sc cutoffpl} &  {\sc cutoffpl $\times$ gabs} & {\sc cutoffpl $\times$ gabs $\times$ gabs} \\
  \hline
 ${\rm const}_{\rm FPMA}$ & 1.000 (frozen)  &  1.000 (frozen) &  1.000 (frozen)  \\ 
 ${\rm const}_{\rm FPMB}$ & $0.982 \pm 0.009$ &  $0.981 \pm 0.009$ &  $0.981 \pm 0.009$ \\ 
 $N_{\rm H}$, $10^{22}$~cm$^{-2}$ & $8.6\pm0.9$ & $5\pm1$ & $2\pm1$ \\
 $E_{\rm fold}$,~keV & $20\pm1$ & $16\pm1$ & $15\pm2$\\ 
 $\Gamma$ & $0.85\pm0.05$ & $0.58^{+0.08}_{-0.09}$ & $0.34^{+0.14}_{-0.18}$   \\ 
 ${\rm Norm}_{\rm continuum}$, ph~keV$^{-1}$~s$^{-1}$~cm$^{-2}$  & $(1.4\pm0.1)\times10^{-2}$ &  $(9\pm1)\times10^{-3}$ & $(6\pm1)\times10^{-3}$   \\
 
 $E_{\rm gabs1}$,~keV &  & $20.4^{+0.7}_{-0.6}$ & $21.0^{+0.7}_{-0.8}$  \\
 $\sigma_{\rm gabs1}$,~keV & & $4\pm1$ & $6^{+2}_{-1}$ \\
 $\tau_{\rm gabs1}$ & &$0.21^{+0.09}_{-0.06}$& $0.38^{+0.29}_{-0.14}$ \\
 $E_{\rm gabs2}$,~keV &  &  &$37\pm1$  \\
 $\sigma_{\rm gabs2}$,~keV & & &$4.5^{+1.5}_{-1.0}$  \\
 $\tau_{\rm gabs2}$ & & & $0.31^{+0.13}_{-0.10}$ \\
 $E_{\rm gauss}$,~keV & 6.46 (frozen) &  6.46 (frozen) & 6.46 (frozen) \\
 $\sigma_{\rm gauss}$,~keV & 0.39 (frozen) &  0.39 (frozen) & 0.39 (frozen) \\
 ${\rm Norm}_{\rm gauss}$, ph~s$^{-1}$~cm$^{-2}$  & $(1.7 \pm 0.5)\times 10^{-4}$& $(2.4 \pm 0.5)\times 10^{-4}$ & $(2.5 \pm 0.5)\times 10^{-4}$ \\
  W-statistic/d.o.f. & 487/361 & 410/358 & 391/355 \\
 \hline
\end{tabular}
\label{table:cyclotron}
\end{center}
\end{table*}

\subsection{Optical and near-IR identification}

\begin{figure}
    \centering
 	\includegraphics[width=0.95\columnwidth]{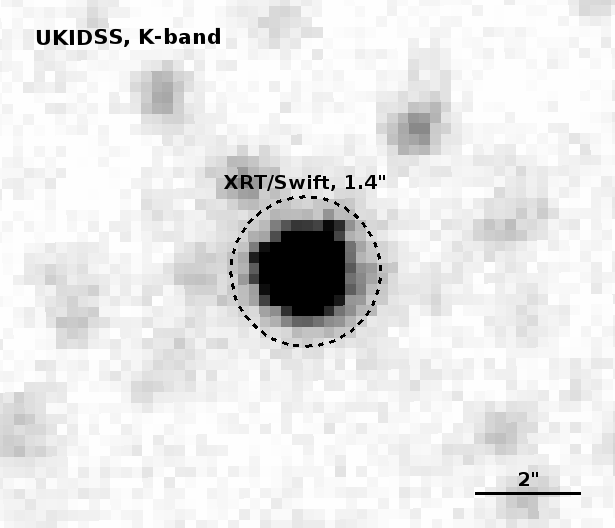} 
	\caption{Image of the sky region containing \sw\ in the IR filter $K$ according to the UKIDSS sky survey (UKIRT telescope). The dashed circle shows the source localization with the {\it Swift}/XRT telescope.}
	\label{fig:UKIDSS}
\end{figure}

First of all, to be sure that the IR-counterpart proposed by \citet{Krimm2014_2} is the only possible for the studied source, we checked the sky region around \sw\ using the GPS/UKIDSS data\footnote{\url{http://wsa.roe.ac.uk/}}. It is essential since UKIDSS has a better angular resolution and greater depth than the 2MASS survey\footnote{\url{https://cdsarc.cds.unistra.fr/viz- bin/cat/II/246}}. As shown in Fig.~\ref{fig:UKIDSS} only one near-IR object from the GPS/UKIDSS survey coinciding with the 2MASS~18082507$-$1753482 star is located inside the error box of {\it Swift}/XRT. Thus, the subsequent analysis was performed for this object. Its magnitudes derived from UKIDSS data are summarized in Table \ref{tab:SWIFT_IR}. As it was mentioned in the Introduction, this star was not found in the DSS survey, that can indicate a quite significant absorption in its direction or inside the binary system. Only the PanSTARRS survey data\footnote{\url{https://panstarrs.stsci.edu}} allowed us to obtain magnitudes and upper limits of the companion in the optical band (Table \ref{tab:SWIFT_IR}). This object is too faint to obtain its optical spectrum with a high S/N ratio, therefore to determine a possible type and class of the companion star we performed a near-IR spectroscopy with NOTCam at the NOT.  

\begin{table}
	\centering
	\caption{Coordinates and magnitudes of the counterpart of \sw\ based on {PanSTARRS and UKIDSS/GPS DR11} data.}
	\label{tab:SWIFT_IR}
	\begin{tabular}{lr}
			\hline \hline
	RA & 18$^{\rm h}$08$^{\rm m}$25\fs08 \\ 
	Dec & -17$\degr53\arcmin48\farcs2$ \\ 
	$l$ & 12\fdg1675 \\
	$b$ &  $+$1\fdg0268 \\
	\hline 
	PanSTARRS\\
	$g$ &  $26.503_{\rm uplim}$  \\
	$r$ &  $22.937_{\rm uplim}$  \\
	$i$ &  $20.152\pm0.016$  \\
	$z$ &  $18.506\pm0.003$  \\
	$y$ &  $17.411\pm0.017$  \\
	\hline 
	UKIDSS\\
	$J$ &  $14.728\pm0.003$  \\
	$H$ &  $13.482\pm0.002$  \\
	$K$ & $12.537\pm0.002$ \\
		\hline
	\end{tabular}
\end{table}

\begin{figure}
    \centering
    \includegraphics[width=1.1\columnwidth]{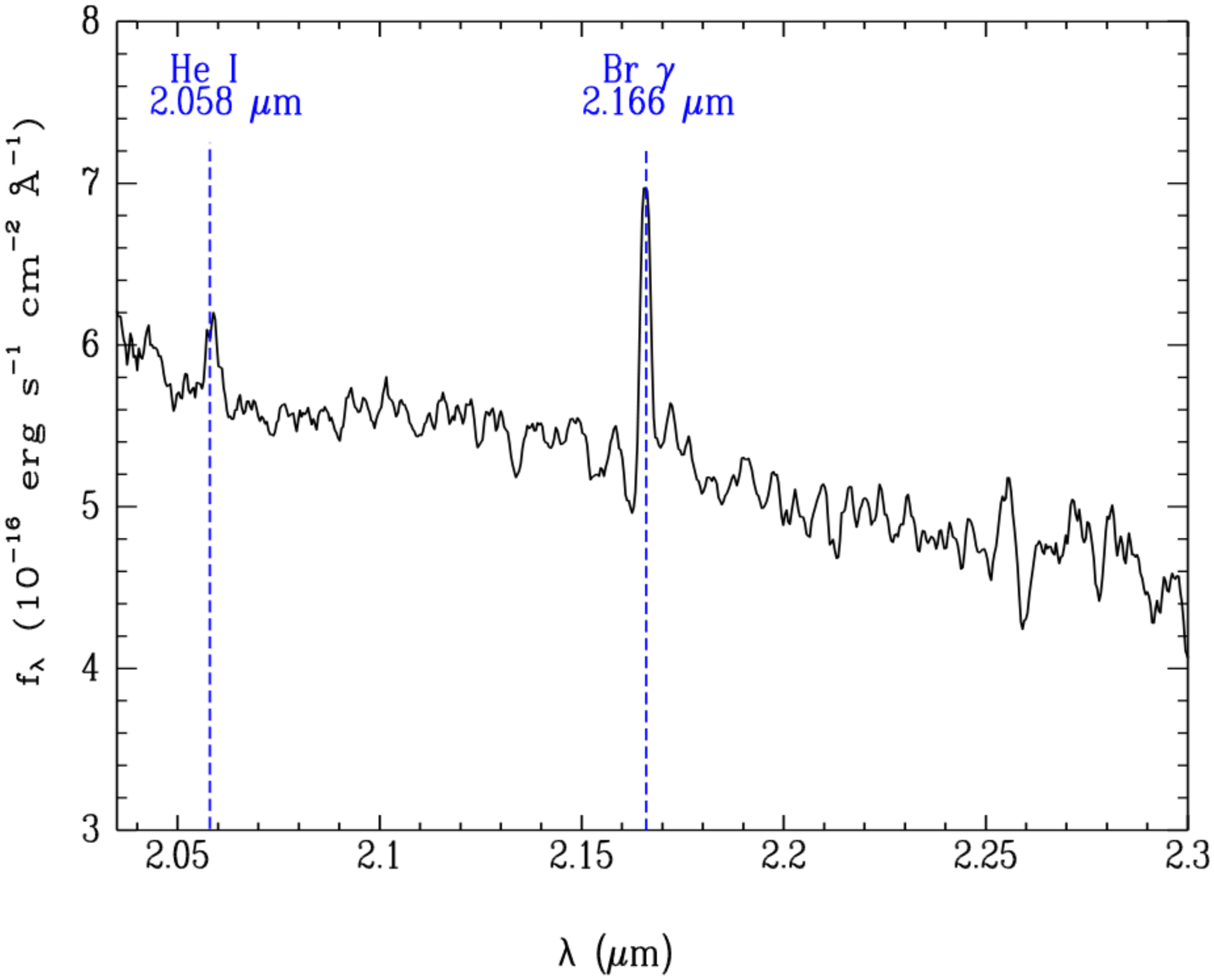}
    \caption{\sw\ $K$-band spectrum obtained with NOTCam on June 20, 2021.}
    \label{fig:kspec}
\end{figure}

The resulting $K$-band spectrum of \sw\ is shown in Fig.~\ref{fig:kspec}. 
We used thirteen 2MASS stars with the photometric quality flag AAA in the acquisition image to calibrate the $K-$band flux, giving a target magnitude of $K$ = 12.3$\pm$0.1 mag, which is 0.3 and 0.24 mag brighter than the 2MASS and UKIDSS magnitudes for \sw. We use this value to approximately flux calibrate the spectrum.
Due to difficult observing conditions with a variable seeing, several of the multiple spectra had to be discarded, and the final spectrum is rather noisy, although smoothed. It shows clearly a narrow and strong 2.166 $\mu$m Brackett~$\gamma$ emission line with an equivalent width of around $-$9 \AA \  and a relatively weak 2.058 $\mu$m He\,I emission line with an EW of about $-$2 \AA \ . The presence of these two lines in emission is typical for Be-type stars \citep{Hanson1996}. Moreover, according to \citet{clark2000} the He\,I 2.058 $\mu$m emission is typically confined to spectral types earlier than B2.5.

\section{Discussion}

\subsection{Distance estimation}

\begin{figure}
    \centering
 	\includegraphics[width=0.95\columnwidth,trim={1.8cm 8.3cm 1cm 4.2cm},clip]{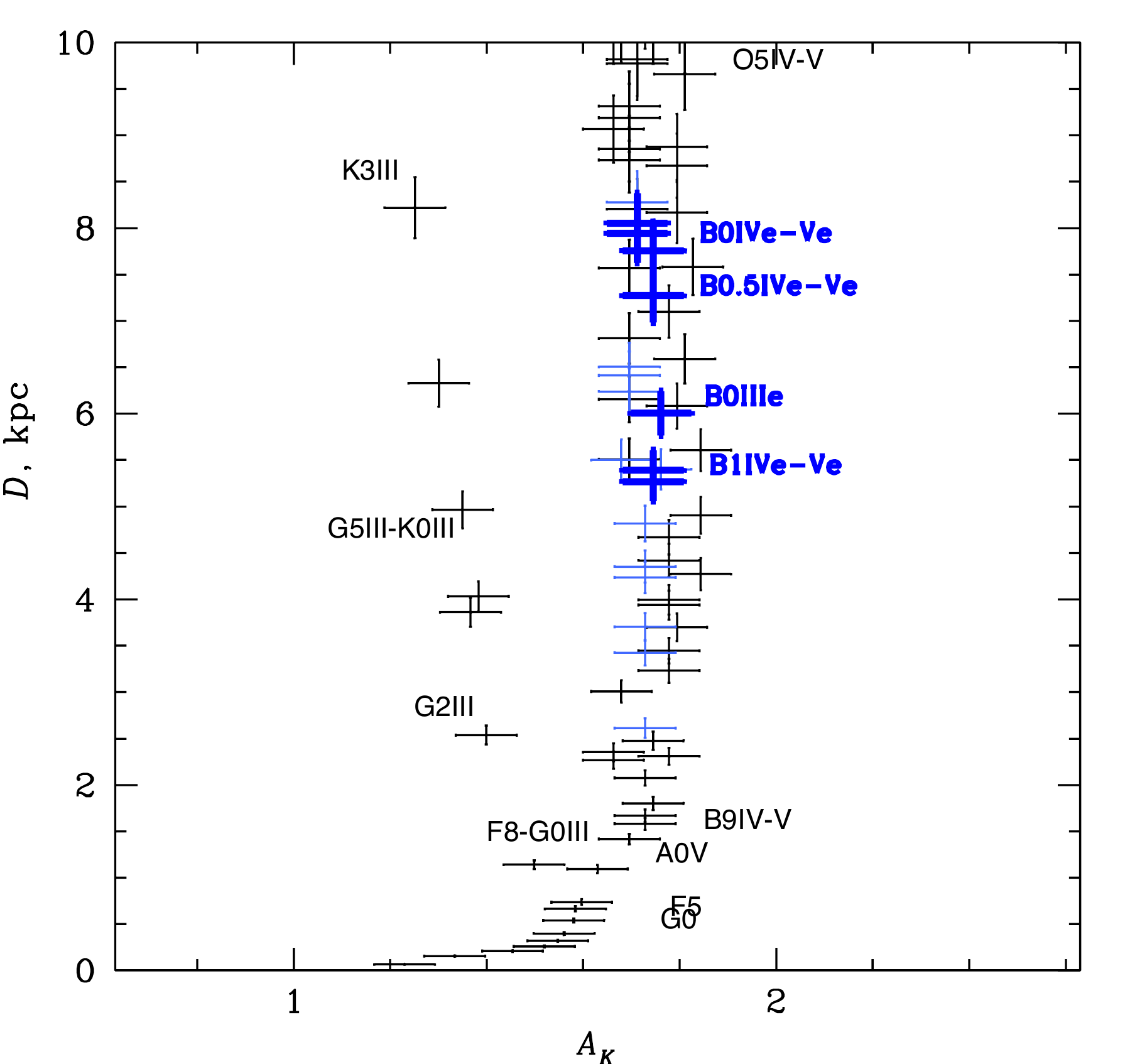}
	\caption{Extinction-Distance plots obtained for possible companion of \sw. It demonstrates at what distance stars of different classes should be located and how strongly they should be absorbed/scattered in order to satisfy the observed magnitudes of near-IR counterpart in the $H$ and $K$ filters from Table~\ref{tab:SWIFT_IR}. Blue crosses mark the subclass of Be-stars. Bold indicates the most probable classes for the IR companion of the source according to the results of $K$-band spectroscopy.}
	\label{fig:IRTYPE}
\end{figure}

To estimate the extinction value in the direction to the IR-star and its distance, we used the method successively applied previously by \citet{Karasev2015, Nabizadeh2019, 2021ApJ...909..154T, Salganik2022} for different sources. By comparing the measured color of the source $(H-K)\simeq0.945$ with intrinsic colors $(H-K)_0$ of various star classes \citet{Wegner2000,Wegner2006,Wegner2007,Wegner2014,Wegner2015}, we can estimate the corresponding reddening $E(H-K)=(H-K)-(H-K)_0$ and extinction $A_{K}$ using a standard extinction law \citep[][]{Cardelli1989}, as \sw\ is located far away from the Galactic bulge direction.
We can also estimate a distance to the source by comparing absolute magnitudes of the same classes of stars $M_{\rm K} $ and the measured magnitude of the source in the $K$-filter using the relation $5-5\log_{10}D=M_{\rm K} - K + A_K$. The result of this approach for different possible types of the companion candidates is shown in Fig.~\ref{fig:IRTYPE}. 

Besides, based on the $K$-band spectroscopy, we know that the counterpart of the source is a Be-star with a spectral type earlier than B2.5. Moreover, looking at samples of Be-stars spectra from \citet{clark2000} we can conclude that B0-B1e is the most likely class of the near-IR counterpart of \sw\ with the corresponding  extinction of $A_{K}\sim1.6-1.8$~mag. Thus, according to Fig.~\ref{fig:IRTYPE} \sw\ should be located as far as $\simeq5-8$~kpc from the Sun. 

\begin{figure}
    \centering
 	\includegraphics[width=0.95\columnwidth,trim={1.3cm 7cm 1cm 3.8cm},clip]{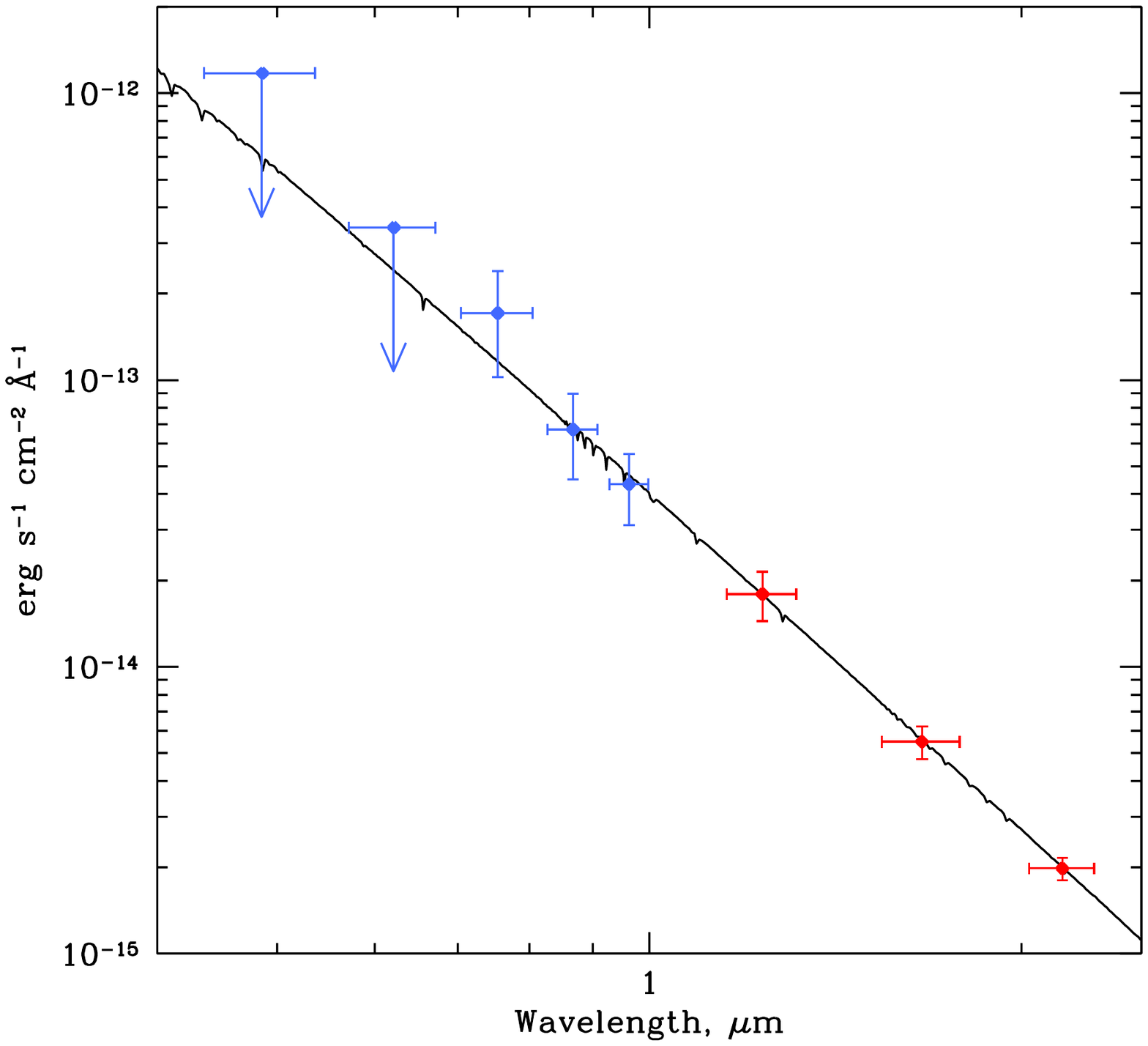}
	\caption{Spectral energy distribution obtained using PanSTARRS (blue crosses) and UKIDSS (red crosses) unabsorbed photometric data. Black line corresponds to the B0-star spectrum taken from the Kurucz 1993 spectral atlas.}
	
	\label{fig:PHOTSED}
\end{figure}

We built a spectral-energy distribution (SED) for the unabsorbed (corrected for the above estimated extinction and distance) magnitudes (and upper limits) of the companion, and compared it with the spectrum of a B0-type star taken from the Kurucz 1993 spectral atlas\footnote{\url{https://www.stsci.edu/hst/instrumentation/reference-data-for-calibration-and-tools/astronomical-catalogs/kurucz-1993-models}} (Fig.~\ref{fig:PHOTSED}). It is clearly seen that the spectrum is in a good agreement with the constructed SED that confirms our estimations. 

According to the \textit{Gaia} data taken from the standard Early Data Release 3 (EDR3) catalog\footnote{\url{https://cdsarc.cds.unistra.fr/viz-bin/cat/I/350}} the source parallax is quite high $4.73\pm1.40$ mas, which corresponds to the distance of $\simeq0.21$ kpc. At the same time, according to the \gaia\ Geometric and photogeometric distances catalog\footnote{\url{https://cdsarc.cds.unistra.fr/viz-bin/cat/I/352}} \citep{Bailer2021}, the geometric distance to the studied object is $1.1^{+0.4}_{-0.6}$ kpc. Both these values are significantly lower in a comparison with our estimates. Moreover, at such distances an early type star like B0-B1e should be much brighter both in optical and IR filters. 

Since the source is significantly absorbed, \gaia\ could not measure its magnitude in the wide $G$-filter, but it was able to obtain the source magnitude in the $G_{\rm RP}$-filter as $18.760\pm0.076$. It should be noted that this magnitude was estimated with a considerable uncertainty as the source is very faint in this band. Moreover, a parameter $visibility\_period\_used$ for this source equals 11, that could be not enough to obtain a precise parallax estimation \footnote{\url{https://gea.esac.esa.int/archive/documentation/GEDR3/Gaia_archive/chap_datamodel/sec_dm_main_tables/ssec_dm_gaia_source.html}}. Thus, due to these limitations and apparent contradictions with our photometric and spectroscopic results, we do not use the \gaia\ data for the distance estimations to \sw.  We expect that new releases will help to clarify the parallax of the source.

The extinction $A_{K}$ can be converted into the hydrogen column density $N_{\rm H}$ using the standard extinction law and the correlation formula $N_{\rm H} = 2.87\times 10^{21} A_V$ \citep{Foight2016}. The obtained value $N_{\rm H}\simeq 3.9-4.4\times 10^{22}$ cm$^{-2}$ agrees well with value of $N_{\rm H}$ derived from the X-ray spectrum (see Table \ref{table:spec_params}).


Another way to estimate the distance to the system is to measure the NS luminosity in the quiescent state and compare it to the expected one. Therefore, we triggered several TOO observations of the source in Aug 22 -- Sep 4, 2021 with a total exposure of 6~ks. As a result we were able to detect the source at the 3.5$\sigma$ level for a circular source region with the radius of 20\arcsec\ with 12 counts and a background region with a radius of 170\arcsec\ with 89 counts using \citet{Gehrels1986} table. Renormalisation and subtraction results in 10.8 counts from the source (corresponding count rate is $2\times10^{-3}$~counts~s$^{-1}$).

The flux-to-count rate ratio obtained in the last observation before the source going to the low state during the 2014 outburst (ObsID 00080746001) was used to convert the count rate measured during the 2021 quiescent state to the flux.
The quiescent flux thus obtained is equal to $2.2\times10^{-13}$ \ergscm, which corresponds to the luminosity of $(0.7-2.0)\times10^{33}$ \ergs\ for the distance of 5-8~kpc. This estimate agrees well with 
the luminosities measured for a sample of Be/XRPs in the quiescent state \citep{Tsygankov2016, Tsygankov2017}.

\subsection{Cyclotron line discovery}

The most direct way to determine a magnetic field of the NS is to find the cyclotron resonant scattering feature (or by other words a cyclotron absorption line) and and measure its energy \citep[see, e.g.,][and references therein]{Staubert2019}. Our pulse phase-resolved analysis revealed a fundamental CRSF at $\sim$21 keV, therefore, using the formula 
\begin{equation}
B_{12} = \frac{1+z}{11.6}\times E_{\rm cyc} = \frac{1}{\sqrt{1-\frac{2GM_{\rm NS}}{R_{\rm NS}\,c^2 }}}\times \frac{E_{\rm cyc}}{11.6}  
\end{equation}
for canonical NS parameters ($M_{\rm NS} = 1.4M_{\sun}$, $R_{\rm NS}=10$ km) one can estimate a strength of the magnetic field of the NS in \sw\ as $B \simeq 2.4 \times 10^{12}$~G. The discovery of CRSF at phases 0.0-0.2 with a depth an order of magnitude greater than the 3$\sigma$ upper limits measured at all other phases indicates the pulse-phase-transient nature of the cyclotron line.  It makes \sw\ one of only a few XRPs where a pulse phase-transient cyclotron line is found.

It is interesting, that sometimes CRSFs manifest themselves in the temporal properties of XRPs. Indeed, it was found for several sources that pulse profiles are distorted at the energies near the CRSFs: V~0332+53 \citep{Tsygankov2006} and 4U~0115+63 \citep{Tsygankov2007,Ferrigno2011}.
This distortion can be interpreted as an appearance of phase lags in the pulse profiles in the vicinity of the cyclotron energy (see Fig. 2 in \citealt{Tsygankov2007}, Fig. 3 in \citealt{Ferrigno2011} and Fig. 9 in \citealt{Tsygankov2006}).
\citet{Schonherr2014} have shown that an addition of a bulk velocity component to the simulation of the intrinsic beam patterns of the redistributed photons by the cyclotron resonant scattering results in the appearance of the energy dependent phase lags similar to reported from observations.

In the case of \sw, the energy-resolved pulse profiles exhibit a similar behavior (see Fig.~\ref{fig:2D}). It also shows the described "wavy" behavior of the phase lags with the largest value reached near the 21 keV. Indeed, near the cyclotron line energy, the pulse profile demonstrates the strongest deformation, leading to a significant decrease of the pulsed fraction (see Fig. ~\ref{fig:pulsed}) and a significant value of the phase lag.  Similar features in the pulsed fraction behaviour near the CRSF energy were also observed in some XRPs \citep[see, e.g.,][]{Tsygankov2007,Ferrigno2009,LutovinovTsygankov2009,2010MNRAS.401.1628T,2017MNRAS.466..593L}. 
Thus, temporal properties of \sw\ can be considered as an independent confirmation for our discovery of the CRSF in the source spectrum.

\section{Summary}

In this paper we provide the first extensive study of XRP \sw\ and its IR companion. The study of \sw\ during the 2014 outburst in a wide energy range of 0.8-79~keV was carried out using data from the \textit{NuSTAR} and \textit{Swift} observatories. The X-ray spectrum of \sw\ has a typical form for XRPs and can be described by a power law with an exponential cutoff. Timing analysis revealed pulsations with a period of 909.73~s and the energy dependent phase lags. The pulsed fraction is increased non-monotonically with the increasing energy, having a local minimum in the range of 17-22 keV. The phase-resolved spectroscopy revealed the presence of CRSF at the pulse phases of 0.0-0.2 with a fundamental energy of 21~keV, which made it possible to estimate the magnetic field of the NS as $B \simeq 2.4\times10^{12}$~G. We attribute the presence of a local minimum in the pulsed fraction and energy-dependent phase lags in the light curves to the influence of CRSF. It is worth noting that phase lags and CRSF were discovered at the same spin phases confirming their physical relation.

We have also studied the companion star of \sw\ based on near-IR spectroscopy obtained with the Nordic Optical Telescope and established that the companion should to be a Be star of a B0-B1 class. The additional use of the UKIDSS catalog allowed us to estimate the distance to the system as  5-8~kpc from the Sun. 

\section*{Acknowledgements}
We are grateful to the \textit{Swift} team for approving and rapid scheduling of the TOO observations. This research has made use of data and software provided by the High Energy Astrophysics Science Archive Research Center (HEASARC), which is a service of the Astrophysics Science Division at NASA/GSFC and the High Energy Astrophysics Division of the Smithsonian Astrophysical Observatory. It also made use of data supplied by the UK \textit{Swift} Science Data Centre at the University of Leicester and data obtained with \textit{NuSTAR} mission, a project led by Caltech, funded by NASA and managed by JPL. This research also has made use of the NuSTAR Data Analysis Software ({\sc NUSTARDAS}) jointly developed by the ASI Science Data Center (ASDC, Italy) and Caltech. This work was supported by the grant 19-12-00423 of the Russian Science Foundation. SST also acknowledge the support from the Academy of Finland travel grants 349373. This work is partly based on observations made with the Nordic Optical Telescope, owned in collaboration by the University of Turku and Aarhus University, and operated jointly by Aarhus University, the University of Turku and the University of Oslo, representing Denmark, Finland and Norway, the University of Iceland and Stockholm University at the Observatorio del Roque de los Muchachos, La Palma, Spain, of the Instituto de Astrofisica de Canarias.

\section*{Data Availability}
{\it NuSTAR} and {\it Swift} data can be accessed from corresponding online archives.
The optical and IR data underlying this article will be shared on reasonable request to the corresponding author.

\bibliographystyle{mnras}
\bibliography{allbib}

\begin{thebibliography}{}
\makeatletter
\relax
\def\mn@urlcharsother{\let\do\@makeother \do\$\do\&\do\#\do\^\do\_\do\%\do\~}
\def\mn@doi{\begingroup\mn@urlcharsother \@ifnextchar [ {\mn@doi@}
  {\mn@doi@[]}}
\def\mn@doi@[#1]#2{\def\@tempa{#1}\ifx\@tempa\@empty \href
  {http://dx.doi.org/#2} {doi:#2}\else \href {http://dx.doi.org/#2} {#1}\fi
  \endgroup}
\def\mn@eprint#1#2{\mn@eprint@#1:#2::\@nil}
\def\mn@eprint@arXiv#1{\href {http://arxiv.org/abs/#1} {{\tt arXiv:#1}}}
\def\mn@eprint@dblp#1{\href {http://dblp.uni-trier.de/rec/bibtex/#1.xml}
  {dblp:#1}}
\def\mn@eprint@#1:#2:#3:#4\@nil{\def\@tempa {#1}\def\@tempb {#2}\def\@tempc
  {#3}\ifx \@tempc \@empty \let \@tempc \@tempb \let \@tempb \@tempa \fi \ifx
  \@tempb \@empty \def\@tempb {arXiv}\fi \@ifundefined
  {mn@eprint@\@tempb}{\@tempb:\@tempc}{\expandafter \expandafter \csname
  mn@eprint@\@tempb\endcsname \expandafter{\@tempc}}}

\bibitem[\protect\citeauthoryear{{Abbott} et~al.,}{{Abbott}
  et~al.}{2000}]{Abbott2000}
{Abbott} T.~M.,  et~al., 2000, in {Iye} M.,  {Moorwood} A.~F.,  eds,  Proc.
  SPIE Vol. 4008, Optical and IR Telescope Instrumentation and Detectors. pp
  714--719, \mn@doi{10.1117/12.395528}

\bibitem[\protect\citeauthoryear{{Arnaud}}{{Arnaud}}{1996}]{Arnaud1996}
{Arnaud} K.~A.,  1996, in {Jacoby} G.~H.,  {Barnes} J.,  eds,  Astronomical
  Society of the Pacific Conference Series Vol. 101, Astronomical Data Analysis
  Software and Systems V. p.~17

\bibitem[\protect\citeauthoryear{{Bachetti}, {Tomsick}  \& {Foster}}{{Bachetti}
  et~al.}{2014}]{Bachetti2014}
{Bachetti} M.,  {Tomsick} J.~A.,   {Foster} K.,  2014, The Astronomer's
  Telegram, \href {https://ui.adsabs.harvard.edu/abs/2014ATel.6229....1B}
  {6229, 1}

\bibitem[\protect\citeauthoryear{{Bailer-Jones}, {Rybizki}, {Fouesneau},
  {Demleitner}  \& {Andrae}}{{Bailer-Jones} et~al.}{2021}]{Bailer2021}
{Bailer-Jones} C.~A.~L.,  {Rybizki} J.,  {Fouesneau} M.,  {Demleitner} M.,
  {Andrae} R.,  2021, \mn@doi [\aj] {10.3847/1538-3881/abd806}, \href
  {https://ui.adsabs.harvard.edu/abs/2021AJ....161..147B} {161, 147}

\bibitem[\protect\citeauthoryear{{Boldin}, {Tsygankov}  \&
  {Lutovinov}}{{Boldin} et~al.}{2013}]{Boldin2013}
{Boldin} P.~A.,  {Tsygankov} S.~S.,   {Lutovinov} A.~A.,  2013, \mn@doi
  [Astronomy Letters] {10.1134/S1063773713060029}, \href
  {https://ui.adsabs.harvard.edu/abs/2013AstL...39..375B} {39, 375}

\bibitem[\protect\citeauthoryear{{Cardelli}, {Clayton}  \& {Mathis}}{{Cardelli}
  et~al.}{1989}]{Cardelli1989}
{Cardelli} J.~A.,  {Clayton} G.~C.,   {Mathis} J.~S.,  1989, \mn@doi [\apj]
  {10.1086/167900}, \href
  {https://ui.adsabs.harvard.edu/abs/1989ApJ...345..245C} {345, 245}

\bibitem[\protect\citeauthoryear{{Clark} \& {Steele}}{{Clark} \&
  {Steele}}{2000}]{clark2000}
{Clark} J.~S.,  {Steele} I.~A.,  2000, \mn@doi [\aaps] {10.1051/aas:2000310},
  \href {https://ui.adsabs.harvard.edu/abs/2000A&AS..141...65C} {141, 65}

\bibitem[\protect\citeauthoryear{{Coburn}, {Heindl}, {Rothschild}, {Gruber},
  {Kreykenbohm}, {Wilms}, {Kretschmar}  \& {Staubert}}{{Coburn}
  et~al.}{2002}]{Coburn2002}
{Coburn} W.,  {Heindl} W.~A.,  {Rothschild} R.~E.,  {Gruber} D.~E.,
  {Kreykenbohm} I.,  {Wilms} J.,  {Kretschmar} P.,   {Staubert} R.,  2002,
  \mn@doi [\apj] {10.1086/343033}, \href
  {https://ui.adsabs.harvard.edu/abs/2002ApJ...580..394C} {580, 394}

\bibitem[\protect\citeauthoryear{{Djupvik} \& {Andersen}}{{Djupvik} \&
  {Andersen}}{2010}]{Djupvik2010}
{Djupvik} A.~A.,  {Andersen} J.,  2010, in Diego J.~M.,  Goicoechea L.~J.,
  Gonz{\'a}lez-Serrano J.~I.,   Gorgas J.,  eds,  Astrophysics and Space
  Science Proceedings Vol. 14, Highlights of Spanish Astrophysics V. Springer,
  Berlin, Heidelberg, p.~211 (\mn@eprint {arXiv} {0901.4015}),
  \mn@doi{10.1007/978-3-642-11250-8\_21}

\bibitem[\protect\citeauthoryear{{Doroshenko}, {Santangelo}, {Doroshenko}  \&
  {Piraino}}{{Doroshenko} et~al.}{2017}]{Doroshenko2017}
{Doroshenko} R.,  {Santangelo} A.,  {Doroshenko} V.,   {Piraino} S.,  2017,
  \mn@doi [\aap] {10.1051/0004-6361/201630137}, \href
  {https://ui.adsabs.harvard.edu/abs/2017A&A...600A..52D} {600, A52}

\bibitem[\protect\citeauthoryear{{Evans} et~al.,}{{Evans}
  et~al.}{2009}]{Evans2009}
{Evans} P.~A.,  et~al., 2009, \mn@doi [\mnras]
  {10.1111/j.1365-2966.2009.14913.x}, \href
  {https://ui.adsabs.harvard.edu/abs/2009MNRAS.397.1177E} {397, 1177}

\bibitem[\protect\citeauthoryear{{Ferrigno}, {Becker}, {Segreto}, {Mineo}  \&
  {Santangelo}}{{Ferrigno} et~al.}{2009}]{Ferrigno2009}
{Ferrigno} C.,  {Becker} P.~A.,  {Segreto} A.,  {Mineo} T.,   {Santangelo} A.,
  2009, \mn@doi [\aap] {10.1051/0004-6361/200809373}, \href
  {https://ui.adsabs.harvard.edu/abs/2009A&A...498..825F} {498, 825}

\bibitem[\protect\citeauthoryear{{Ferrigno}, {Falanga}, {Bozzo}, {Becker},
  {Klochkov}  \& {Santangelo}}{{Ferrigno} et~al.}{2011}]{Ferrigno2011}
{Ferrigno} C.,  {Falanga} M.,  {Bozzo} E.,  {Becker} P.~A.,  {Klochkov} D.,
  {Santangelo} A.,  2011, \mn@doi [\aap] {10.1051/0004-6361/201116826}, \href
  {https://ui.adsabs.harvard.edu/abs/2011A&A...532A..76F} {532, A76}

\bibitem[\protect\citeauthoryear{{Filippova}, {Tsygankov}, {Lutovinov}  \&
  {Sunyaev}}{{Filippova} et~al.}{2005}]{Filippova2005}
{Filippova} E.~V.,  {Tsygankov} S.~S.,  {Lutovinov} A.~A.,   {Sunyaev} R.~A.,
  2005, \mn@doi [Astronomy Letters] {10.1134/1.2123288}, \href
  {https://ui.adsabs.harvard.edu/abs/2005AstL...31..729F} {31, 729}

\bibitem[\protect\citeauthoryear{{Foight}, {G{\"u}ver}, {{\"O}zel}  \&
  {Slane}}{{Foight} et~al.}{2016}]{Foight2016}
{Foight} D.~R.,  {G{\"u}ver} T.,  {{\"O}zel} F.,   {Slane} P.~O.,  2016,
  \mn@doi [\apj] {10.3847/0004-637X/826/1/66}, \href
  {https://ui.adsabs.harvard.edu/abs/2016ApJ...826...66F} {826, 66}

\bibitem[\protect\citeauthoryear{{Gehrels}}{{Gehrels}}{1986}]{Gehrels1986}
{Gehrels} N.,  1986, \mn@doi [\apj] {10.1086/164079}, \href
  {https://ui.adsabs.harvard.edu/abs/1986ApJ...303..336G} {303, 336}

\bibitem[\protect\citeauthoryear{{HI4PI Collaboration} et~al.,}{{HI4PI
  Collaboration} et~al.}{2016}]{HI4PI2016}
{HI4PI Collaboration} et~al., 2016, \mn@doi [\aap]
  {10.1051/0004-6361/201629178}, \href
  {https://ui.adsabs.harvard.edu/abs/2016A&A...594A.116H} {594, A116}

\bibitem[\protect\citeauthoryear{{Hanson}, {Conti}  \& {Rieke}}{{Hanson}
  et~al.}{1996}]{Hanson1996}
{Hanson} M.~M.,  {Conti} P.~S.,   {Rieke} M.~J.,  1996, \mn@doi [\apjs]
  {10.1086/192366}, \href
  {https://ui.adsabs.harvard.edu/abs/1996ApJS..107..281H} {107, 281}

\bibitem[\protect\citeauthoryear{{Harrison} et~al.,}{{Harrison}
  et~al.}{2013}]{Harrison2013}
{Harrison} F.~A.,  et~al., 2013, \mn@doi [\apj] {10.1088/0004-637X/770/2/103},
  \href {https://ui.adsabs.harvard.edu/abs/2013ApJ...770..103H} {770, 103}

\bibitem[\protect\citeauthoryear{{Kaastra} \& {Bleeker}}{{Kaastra} \&
  {Bleeker}}{2016}]{Kaastra2016}
{Kaastra} J.~S.,  {Bleeker} J.~A.~M.,  2016, \mn@doi [\aap]
  {10.1051/0004-6361/201527395}, \href
  {https://ui.adsabs.harvard.edu/abs/2016A&A...587A.151K} {587, A151}

\bibitem[\protect\citeauthoryear{{Karasev}, {Tsygankov}  \&
  {Lutovinov}}{{Karasev} et~al.}{2015}]{Karasev2015}
{Karasev} D.~I.,  {Tsygankov} S.~S.,   {Lutovinov} A.~A.,  2015, \mn@doi
  [Astronomy Letters] {10.1134/S1063773715080022}, \href
  {https://ui.adsabs.harvard.edu/abs/2015AstL...41..394K} {41, 394}

\bibitem[\protect\citeauthoryear{{Kreykenbohm}, {Coburn}, {Wilms},
  {Kretschmar}, {Staubert}, {Heindl}  \& {Rothschild}}{{Kreykenbohm}
  et~al.}{2002}]{Kreykenbohm2002}
{Kreykenbohm} I.,  {Coburn} W.,  {Wilms} J.,  {Kretschmar} P.,  {Staubert} R.,
  {Heindl} W.~A.,   {Rothschild} R.~E.,  2002, \mn@doi [\aap]
  {10.1051/0004-6361:20021181}, \href
  {https://ui.adsabs.harvard.edu/abs/2002A&A...395..129K} {395, 129}

\bibitem[\protect\citeauthoryear{{Krimm} et~al.,}{{Krimm}
  et~al.}{2014a}]{Krimm2014}
{Krimm} H.~A.,  et~al., 2014a, The Astronomer's Telegram, \href
  {https://ui.adsabs.harvard.edu/abs/2014ATel.6138....1K} {6138, 1}

\bibitem[\protect\citeauthoryear{{Krimm}, {Kennea}  \& {Holland}}{{Krimm}
  et~al.}{2014b}]{Krimm2014_2}
{Krimm} H.~A.,  {Kennea} J.~A.,   {Holland} S.~T.,  2014b, The Astronomer's
  Telegram, \href {https://ui.adsabs.harvard.edu/abs/2014ATel.6155....1K}
  {6155, 1}

\bibitem[\protect\citeauthoryear{{Lutovinov} \& {Tsygankov}}{{Lutovinov} \&
  {Tsygankov}}{2009}]{LutovinovTsygankov2009}
{Lutovinov} A.~A.,  {Tsygankov} S.~S.,  2009, \mn@doi [Astronomy Letters]
  {10.1134/S1063773709070019}, \href
  {https://ui.adsabs.harvard.edu/abs/2009AstL...35..433L} {35, 433}

\bibitem[\protect\citeauthoryear{{Lutovinov}, {Tsygankov}, {Postnov},
  {Krivonos}, {Molkov}  \& {Tomsick}}{{Lutovinov}
  et~al.}{2017}]{2017MNRAS.466..593L}
{Lutovinov} A.~A.,  {Tsygankov} S.~S.,  {Postnov} K.~A.,  {Krivonos} R.~A.,
  {Molkov} S.~V.,   {Tomsick} J.~A.,  2017, \mn@doi [\mnras]
  {10.1093/mnras/stw3058}, \href
  {https://ui.adsabs.harvard.edu/abs/2017MNRAS.466..593L} {466, 593}

\bibitem[\protect\citeauthoryear{{Molkov}, {Lutovinov}, {Tsygankov},
  {Mereminskiy}  \& {Mushtukov}}{{Molkov} et~al.}{2019}]{Molkov2019}
{Molkov} S.,  {Lutovinov} A.,  {Tsygankov} S.,  {Mereminskiy} I.,   {Mushtukov}
  A.,  2019, \mn@doi [\apjl] {10.3847/2041-8213/ab3e4d}, \href
  {https://ui.adsabs.harvard.edu/abs/2019ApJ...883L..11M} {883, L11}

\bibitem[\protect\citeauthoryear{{Molkov}, {Doroshenko}, {Lutovinov},
  {Tsygankov}, {Santangelo}, {Mereminskiy}  \& {Semena}}{{Molkov}
  et~al.}{2021}]{Molkov2021}
{Molkov} S.,  {Doroshenko} V.,  {Lutovinov} A.,  {Tsygankov} S.,  {Santangelo}
  A.,  {Mereminskiy} I.,   {Semena} A.,  2021, \mn@doi [\apjl]
  {10.3847/2041-8213/ac0c15}, \href
  {https://ui.adsabs.harvard.edu/abs/2021ApJ...915L..27M} {915, L27}

\bibitem[\protect\citeauthoryear{{Mushtukov} \& {Tsygankov}}{{Mushtukov} \&
  {Tsygankov}}{2022}]{2022arXiv220414185M}
{Mushtukov} A.,  {Tsygankov} S.,  2022, arXiv e-prints, \href
  {https://ui.adsabs.harvard.edu/abs/2022arXiv220414185M} {p. arXiv:2204.14185}

\bibitem[\protect\citeauthoryear{{Nabizadeh}, {Tsygankov}, {Karasev},
  {M{\"o}nkk{\"o}nen}, {Lutovinov}, {Nagirner}  \& {Poutanen}}{{Nabizadeh}
  et~al.}{2019}]{Nabizadeh2019}
{Nabizadeh} A.,  {Tsygankov} S.~S.,  {Karasev} D.~I.,  {M{\"o}nkk{\"o}nen} J.,
  {Lutovinov} A.~A.,  {Nagirner} D.~I.,   {Poutanen} J.,  2019, \mn@doi [\aap]
  {10.1051/0004-6361/201834635}, \href
  {https://ui.adsabs.harvard.edu/abs/2019A&A...622A.198N} {622, A198}

\bibitem[\protect\citeauthoryear{{Romano} et~al.,}{{Romano}
  et~al.}{2014}]{Romano2014}
{Romano} P.,  et~al., 2014, \mn@doi [\aap] {10.1051/0004-6361/201322516}, \href
  {https://ui.adsabs.harvard.edu/abs/2014A&A...562A...2R} {562, A2}

\bibitem[\protect\citeauthoryear{{Salganik}, {Tsygankov}, {Djupvik}, {Karasev},
  {Lutovinov}, {Buckley}, {Gromadzki}  \& {Poutanen}}{{Salganik}
  et~al.}{2022}]{Salganik2022}
{Salganik} A.,  {Tsygankov} S.~S.,  {Djupvik} A.~A.,  {Karasev} D.~I.,
  {Lutovinov} A.~A.,  {Buckley} D. A.~H.,  {Gromadzki} M.,   {Poutanen} J.,
  2022, \mn@doi [\mnras] {10.1093/mnras/stab3362}, \href
  {https://ui.adsabs.harvard.edu/abs/2022MNRAS.509.5955S} {509, 5955}

\bibitem[\protect\citeauthoryear{{Sch{\"o}nherr} et~al.,}{{Sch{\"o}nherr}
  et~al.}{2014}]{Schonherr2014}
{Sch{\"o}nherr} G.,  et~al., 2014, \mn@doi [\aap]
  {10.1051/0004-6361/201322448}, \href
  {https://ui.adsabs.harvard.edu/abs/2014A&A...564L...8S} {564, L8}

\bibitem[\protect\citeauthoryear{{Staubert} et~al.,}{{Staubert}
  et~al.}{2019}]{Staubert2019}
{Staubert} R.,  et~al., 2019, \mn@doi [\aap] {10.1051/0004-6361/201834479},
  \href {https://ui.adsabs.harvard.edu/abs/2019A&A...622A..61S} {622, A61}

\bibitem[\protect\citeauthoryear{{Titarchuk}}{{Titarchuk}}{1994}]{Titarchuk1994}
{Titarchuk} L.,  1994, \mn@doi [\apj] {10.1086/174760}, \href
  {https://ui.adsabs.harvard.edu/abs/1994ApJ...434..570T} {434, 570}

\bibitem[\protect\citeauthoryear{{Tomsick}, {Bachetti}, {Kennea}  \&
  {Krimm}}{{Tomsick} et~al.}{2014}]{Tomsick2014}
{Tomsick} J.~A.,  {Bachetti} M.,  {Kennea} J.~A.,   {Krimm} H.~A.,  2014, The
  Astronomer's Telegram, \href
  {https://ui.adsabs.harvard.edu/abs/2014ATel.6170....1T} {6170, 1}

\bibitem[\protect\citeauthoryear{{Tsygankov}, {Lutovinov}, {Churazov}  \&
  {Sunyaev}}{{Tsygankov} et~al.}{2006}]{Tsygankov2006}
{Tsygankov} S.~S.,  {Lutovinov} A.~A.,  {Churazov} E.~M.,   {Sunyaev} R.~A.,
  2006, \mn@doi [\mnras] {10.1111/j.1365-2966.2006.10610.x}, \href
  {https://ui.adsabs.harvard.edu/abs/2006MNRAS.371...19T} {371, 19}

\bibitem[\protect\citeauthoryear{{Tsygankov}, {Lutovinov}, {Churazov}  \&
  {Sunyaev}}{{Tsygankov} et~al.}{2007}]{Tsygankov2007}
{Tsygankov} S.~S.,  {Lutovinov} A.~A.,  {Churazov} E.~M.,   {Sunyaev} R.~A.,
  2007, \mn@doi [Astronomy Letters] {10.1134/S1063773707060023}, \href
  {https://ui.adsabs.harvard.edu/abs/2007AstL...33..368T} {33, 368}

\bibitem[\protect\citeauthoryear{{Tsygankov}, {Lutovinov}  \&
  {Serber}}{{Tsygankov} et~al.}{2010}]{2010MNRAS.401.1628T}
{Tsygankov} S.~S.,  {Lutovinov} A.~A.,   {Serber} A.~V.,  2010, \mn@doi
  [\mnras] {10.1111/j.1365-2966.2009.15791.x}, \href
  {https://ui.adsabs.harvard.edu/abs/2010MNRAS.401.1628T} {401, 1628}

\bibitem[\protect\citeauthoryear{{Tsygankov}, {Lutovinov}, {Doroshenko},
  {Mushtukov}, {Suleimanov}  \& {Poutanen}}{{Tsygankov}
  et~al.}{2016}]{Tsygankov2016}
{Tsygankov} S.~S.,  {Lutovinov} A.~A.,  {Doroshenko} V.,  {Mushtukov} A.~A.,
  {Suleimanov} V.,   {Poutanen} J.,  2016, \mn@doi [\aap]
  {10.1051/0004-6361/201628236}, \href
  {https://ui.adsabs.harvard.edu/abs/2016A&A...593A..16T} {593, A16}

\bibitem[\protect\citeauthoryear{{Tsygankov}, {Wijnands}, {Lutovinov},
  {Degenaar}  \& {Poutanen}}{{Tsygankov} et~al.}{2017}]{Tsygankov2017}
{Tsygankov} S.~S.,  {Wijnands} R.,  {Lutovinov} A.~A.,  {Degenaar} N.,
  {Poutanen} J.,  2017, \mn@doi [\mnras] {10.1093/mnras/stx1255}, \href
  {https://ui.adsabs.harvard.edu/abs/2017MNRAS.470..126T} {470, 126}

\bibitem[\protect\citeauthoryear{{Tsygankov} et~al.,}{{Tsygankov}
  et~al.}{2021}]{2021ApJ...909..154T}
{Tsygankov} S.~S.,  et~al., 2021, \mn@doi [\apj] {10.3847/1538-4357/abddbd},
  \href {https://ui.adsabs.harvard.edu/abs/2021ApJ...909..154T} {909, 154}

\bibitem[\protect\citeauthoryear{{Wachter}, {Leach}  \& {Kellogg}}{{Wachter}
  et~al.}{1979}]{Wachter1979}
{Wachter} K.,  {Leach} R.,   {Kellogg} E.,  1979, \mn@doi [\apj]
  {10.1086/157084}, \href
  {https://ui.adsabs.harvard.edu/abs/1979ApJ...230..274W} {230, 274}

\bibitem[\protect\citeauthoryear{{Wegner}}{{Wegner}}{2000}]{Wegner2000}
{Wegner} W.,  2000, \mn@doi [\mnras] {10.1046/j.1365-8711.2000.03884.x}, \href
  {https://ui.adsabs.harvard.edu/abs/2000MNRAS.319..771W} {319, 771}

\bibitem[\protect\citeauthoryear{{Wegner}}{{Wegner}}{2006}]{Wegner2006}
{Wegner} W.,  2006, \mn@doi [\mnras] {10.1111/j.1365-2966.2006.10549.x}, \href
  {https://ui.adsabs.harvard.edu/abs/2006MNRAS.371..185W} {371, 185}

\bibitem[\protect\citeauthoryear{{Wegner}}{{Wegner}}{2007}]{Wegner2007}
{Wegner} W.,  2007, \mn@doi [\mnras] {10.1111/j.1365-2966.2006.11265.x}, \href
  {https://ui.adsabs.harvard.edu/abs/2007MNRAS.374.1549W} {374, 1549}

\bibitem[\protect\citeauthoryear{{Wegner}}{{Wegner}}{2014}]{Wegner2014}
{Wegner} W.,  2014, \actaa, \href
  {https://ui.adsabs.harvard.edu/abs/2014AcA....64..261W} {64, 261}

\bibitem[\protect\citeauthoryear{{Wegner}}{{Wegner}}{2015}]{Wegner2015}
{Wegner} W.,  2015, \mn@doi [Astronomische Nachrichten]
  {10.1002/asna.201312143}, \href
  {https://ui.adsabs.harvard.edu/abs/2015AN....336..159W} {336, 159}

\bibitem[\protect\citeauthoryear{{Wilms}, {Allen}  \& {McCray}}{{Wilms}
  et~al.}{2000}]{Wilms2000}
{Wilms} J.,  {Allen} A.,   {McCray} R.,  2000, \mn@doi [\apj] {10.1086/317016},
  \href {https://ui.adsabs.harvard.edu/abs/2000ApJ...542..914W} {542, 914}

\makeatother
\end{thebibliography}

\bsp    
\label{lastpage}
\end{document}